\documentclass[aps,prb,preprint,showpacs,footinbib,longbibliography]{revtex4-1}                                                                                                                                                  

\usepackage{graphicx}
\usepackage{amsmath}
\usepackage{amsbsy}
\usepackage{braket}
\usepackage{amssymb}
\usepackage{dcolumn}
\usepackage{bm}

\begin{document}

\title {Chirality Induced Propagation Velocity Asymmetry}

\author{Diego A. Hoff}
\address{Universidade Federal da Fronteira Sul,
SC, Brazil}
\author{Luis G. C. Rego}
\address{Department of Physics, Universidade Federal de Santa Catarina,
SC, 88040-900, Brazil}

\begin{abstract}
The spin-dependent propagation of electrons in helical nanowires is investigated. We show that the interplay of spin angular momentum and nanowire chirality, 
under spin-orbit interaction, lifts the symmetry between left and right propagating electrons, giving rise to a velocity asymmetry.
The study is based on a microscopic tight-binding model that takes into account the spin-orbit interaction. The continuity equation for the spin-dependent probability
density is derived, including the spin non-conserving terms, and quantum dynamics calculations are performed to obtain the electron propagating dynamics. 
The calculations
are applied to the inorganic double-helix SnIP, a quasi-1D material that constitutes a semiconductor with a band gap of $\sim$ 1.9 eV. 
The results, nevertheless, have general validity due to  symmetry considerations. 
The relation  of the propagation velocity asymmetry with the phenomena ascribed to the chiral-induced spin selectivity (CISS) effect is examined.\\
\end{abstract}

\maketitle
\newpage

\vskip2pc

Geometrical chirality is easily noticed for a figure whose image in a plane mirror cannot be 
brought to coincidence with itself.
However, this definition is limited because it does not include the broader notion of dynamic chirality, which is 
more elusive and has far more reaching consequences. \cite{BarronJACS,Wagniere}
The chiral properties of quasi-1D systems have been known from some time. The first systematic study of quantum chiral properties occurred in the 1980's, with the 
discovery of the Integer Quantum Hall effect --  and later the Fractional effect -- and the explanation thereof in terms of chiral edge states.\cite{prange2012} 
More recently,
quantum chiral properties have been investigated in single wall carbon nanotubes (SWCNT), 
\cite{Dresselhaus,KaneMele,Ando,GuineaPRB,SWCNT-Izumida}
which can exhibit chiral phenomena depending on their wrapping 
configuration.\cite{Wilder1998} 
Interestingly, the oldest, most ubiquitous and well known chiral quasi-1D material occurs naturally, the DNA.\cite{WATSON1953} 
Nevertheless, until very recently, its quantum chiral 
properties have gone unnoticed. In a recent experiment, researchers observed that a double-stranded DNA (dsDNA) could generate a spin-polarized electronic current 
out of a spin-unpolarized electron influx.\cite{CISS-discover,SecondCISS} 
The effect was named chiral induced spin selectivity (CISS).\cite{Waldeck2019} From this initial observation various other manifestations 
of the CISS effect were reported, in a broad range of situations, including separation of chiral enantiomers,\cite{Banerjee-Ghosh} 
spin-filtering,\cite{Mujica2017,Gosh-spinfilter} chiral induced spin-LED devices fabricated with layered perovskites,\cite{Kim2021} 
and improved electrocatalytic water splitting with 
chiral metal-oxide films.\cite{electrocatalysis-1,electrocatalysis-2}
Despite the various experimental reports, the fundamental understanding of the effect is still incomplete.
Besides the nature of the CISS effect, its surprising robustness in face of the smallness of the spin-orbit interaction in organic materials, 
even at ambient conditions, constitutes one of the main challenging questions.
In fact, several studies have reproduced qualitatively
the spin filtering effects observed experimentally, but the calculations generally describe a much weaker effect.
\cite{Cuniberti2013,Cuniberti,GuoPNAS,Hedegard2019,MujicaJCTC,Varela}
Most of the theoretical work is based on the stationary Landauer-B\"uttiker transport formalism,
\cite{Cuniberti2013,Cuniberti,Hedegard2019,MujicaJCTC,CunibertiJPCL,VarelaPRB}
some studies place  emphasis on the role played by the contact leads,\cite{Hedegard2019,vanWees2019,vanWees2020}
while others report that scattering, decoherence and/or leakage processes are necessary to describe the magnitude of the 
effect.\cite{GuoPRL,GuoPNAS,Balseiro,Cuniberti,vanWees2020}
Varela et al.\cite{VarelaPRB} reported that by considering the sequential tunneling between well defined localized states large 
spin polarization could be attained with realistic spin-orbit coupling parameters.

In this letter we study the spin-dependent propagation of electrons in helical nanowires. We show that the symmetry of left- and right-propagating electrons
is broken due to the spin-orbit interaction, so that the mobility of electrons in structures with axial chirality depends on the combination of 3 factors: the
direction of the spin angular momentum, the direction of motion, and the chirality of the underlying physical structure. The study was performed in the totally 
inorganic SnIP double-helix system.\cite{SnIP-1,SnIP-properties} 
We argue that the propagation velocity asymmetry can be one of the underlying effects responsible for the CISS effect.
We also point out that the propagation velocity
asymmetry is also described in carbon nanotubes\cite{SWCNT-Izumida} and chiral metamaterials,\cite{AsymmetricV-PRL} thus, 
putting the CISS effect on a common ground with other phenomena.

\begin{figure}[htbp]
    \centering
 \includegraphics[width=0.5\linewidth]{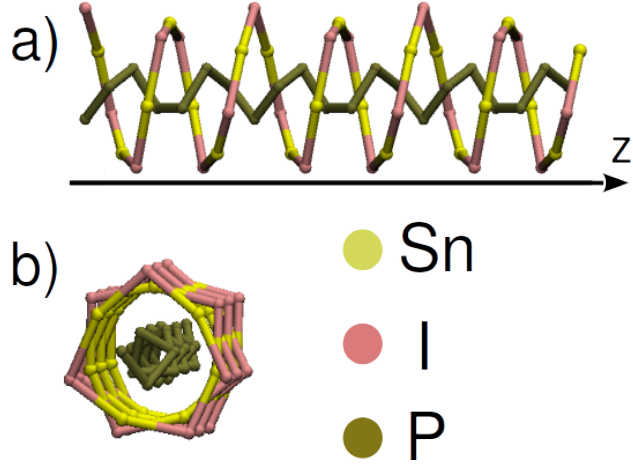}                             
 \caption{Geometrical structure of a left-handed M-SnIP double-helix comprised of 3 unit cells: a) aligned along the $z$ axis and b) frontal viewpoint.
Notice that both the inner [P]$^-$ and the outer [SnI]$^+$ helices have the same left-handed M-chirality; the plus and minus signs designate the oxidation states.}
\label{SnIP}
\end{figure}

{\bf Model system.}
We briefly describe the characteristics of the tin iodide phosphide (SnIP) double-helix,\cite{SnIP-1,SnIP-properties}
shown in Figure \ref{SnIP}, which has a totally inorganic double-helical structure. 
The SnIP double-helix consists of an inner [P]$^-$ (phosphorus) helix wrapped by an outer [SnI]$^+$ (tin iodide) helix, giving rise to a non-magnetic 
double-helix of 0.98 nm in outer diameter. 
We call attention to the fact that both the inner [P]$^-$ and the outer [SnI]$^+$ helices comprising a given SnIP double-helix have the same chirality; 
here the plus and minus signs designate the oxidation states.
Each nanorod is held together by strong intra-helix covalent bonds and a dative ionic interaction between the inner and outer helices.
As synthesized, the nanorods are assembled in bundles by van der Waals forces, giving rise to a semiconductor material that 
exhibits quasi-1D quantum confinement effects.
The bundles consist of a racemic mixture of left-handed and right-handed double-helices aligned parallel to each other in a pseudo-hexagonal arrangement. 
Material characterization shows that the SnIP double-helices constitute a material with an indirect band gap of 1.80 eV and a direct band gap 
of 1.86 eV.\cite{SnIP-1} 
In addition, the material is highly flexible and stable, showing polymer like behavior.\cite{SnIP-properties} 
Time-resolved terahertz (THz) spectroscopy revealed a high intrinsic electron mobility along the double-helix axis, 
which is limited by traps.\cite{purschke2021}
Individual SnIP double-helix strands can be obtained from the racemic mixture by minimizing the van de Waals force and stabilizing the single strands 
in a different medium, such as in SnIP@C$_3$N$_4$(F,Cl) thin films.\cite{SnIP-properties} 
Another possibility, as indicated by total energy calculations,
\cite{SnIP-properties,SnIP-NWs} is the encapsulation of individual strands in carbon nanotubes.  
Herein, we use the geometrical parameters obtained in reference \cite{SnIP-1} to build the unit cell of the model system. 
Within the helices, bond distances vary as: d(Sn-I) = 3.060 to 3.288 \AA, for the outer tin-iodide helix, and d(P-P) = 2.17 to 2.21 \AA~for the inner phosphorus 
helix.  Seven SnIP units comprise the double-helix unit cell,  with a lattice parameter of a = 7.934 \AA. Figure \ref{SnIP} shows a 3 unit cell left-handed strand.

{\bf Theory and Methods.}
The single particle hamiltonian of an electron in the double-helical SnIP nanowire, including the spin-orbit interaction term, is given by
\begin{equation}
H = \frac{1}{2m}\mathbf{p}^2 + V(\mathbf{r}) + 
\frac{1}{2(mc)^2}\mathbf{S}\cdot\boldsymbol{\nabla} V(\mathbf{r}) \times\mathbf{p},
\label{H}
\end{equation}
where $\mathbf{S}=\tfrac{\hbar}{2}\boldsymbol{\sigma}=\tfrac{\hbar}{2}(\sigma_x,\sigma_y,\sigma_z)$ is the spin vector operator comprised of the $\sigma$ 
Pauli matrices, 
$\mathbf{p}=-i\hbar\boldsymbol{\nabla}$ is the electron momentum operator and $V(\mathbf{r})$ is the total
Coulomb potential energy felt by the electron.
We make use of the tight-binding (TB) formalism. 
Several studies have used the TB method, either in the Slater-Koster\cite{GuineaPRB,Medina2016,Hedegard2019,Cuniberti} (SK)
or the extended-H\"uckel\cite{Cerda,KirzcenowMn12} (eH)
framework, to describe carbon nanotubes and  DNA molecules.
Herein, we adopt the eH approach. The total hamiltonian, Eq. (\ref{H}),  is comprised of a spinless part ($H_0$), associated with the eH-TB hamiltonian, 
and the spin-orbit interaction part.
We compute the TB hamiltonian in the basis of atomic Slater-type orbitals (STO's),
$|\chi_{nlm}s\rangle$, 
where $i=\{nlm\}$ represents the orbital quantum numbers associated with  the wavefunction 
$\chi_{nlm}(\mathbf{r}_\alpha)=\mathcal{R}_n(r_\alpha) Y_{lm}(\theta,\varphi)$,
and $s=\{-1,1\}$ (lowercase $s$) designates the spin state of the electron along the $z$ direction. 
The wavefunction components $\mathcal{R}_n(r_\alpha)$ and 
$Y_{lm}(\theta,\varphi)$ describe, respectively, the radial and the angular parts of the STO for an atom located at position $\mathbf{R}_\alpha$, with 
$\mathbf{r}_\alpha = \mathbf{r} - \mathbf{R}_\alpha$.
The basis set includes the 3s and 3p atomic orbitals for the phosphorus (P) atoms and the 5s and 5p orbitals for the tin (Sn) and iodine (I) atoms. 
The matrix elements of the spinless semiempirical eH hamiltonian are provided as Supporting Information.
The spin-orbit (SO) interaction part
\begin{equation}
H_{SO} = \frac{1}{2(mc)^2} \mathbf{S}\cdot\boldsymbol{\nabla} V(\mathbf{r})\times \mathbf{p}
\label{Hso}
\end{equation}
is also written in the framework of the eH-TB formalism.\cite{KirzcenowMn12}
Despite  the complexity of the SO term, the main contribution of the SO interaction arises from the one-electron operators
that couple the electron with the nuclei.\cite{Moores1973PRSL} 
Therefore, in the TB method, the total Coulomb potential can be approximated by a summation over spherically 
symmetric single particle potentials produced by all the nuclei $\alpha$, 
$V(\mathbf{r}) \approx \sum_\alpha V_\alpha(|\mathbf{r}-\mathbf{R}_\alpha|)$, so that 
\begin{equation}
H_{SO} \approx \sum_\alpha \dfrac{1}{2m^2c^2}\dfrac{1}{|\mathbf{r}_\alpha|} 
\dfrac{dV_\alpha(r_\alpha)}{dr_\alpha} \mathbf{S}\cdot\mathbf{L}_\alpha
= \sum_\alpha \hat{\lambda}_\alpha(r_\alpha) \frac{\mathbf{S}\cdot\mathbf{L}_\alpha}{\hbar^2},
\label{SO1}
\end{equation}
with $\mathbf{L}_\alpha = (\mathbf{r} -\mathbf{R}_\alpha)\times \mathbf{p}$ describing the angular momentum of the electron with respect to atom $\alpha$.
It has been shown that the effective single particle SO operator $\xi_\alpha \mathbf{S}\cdot\mathbf{L}_\alpha$ produces very good results\cite{Moores1973PRSL} 
when the coupling parameter $\xi_\alpha$ is obtained from ab-initio calculations or experimental fits, so as to incorporate SO many-body effects.

The matrix elements of $H_{SO}$ can be written as 
\begin{equation}
\langle is_i,\beta|H_{SO}|js_j,\gamma\rangle=\sum_{\alpha}\langle is_i,\beta|\left(\hat{\lambda}_\alpha(r_{\alpha})
\dfrac{\mathbf{S}\cdot\mathbf{L}_\alpha}{\hbar^2}\right)|js_j,\gamma\rangle,
\label{eq:1_2}
\end{equation}
with $\beta$ and $\gamma$ designating the atoms located at positions $\mathbf{R}_\beta$ and $\mathbf{R}_\gamma$. 
The matrix element of Eq. (\ref{eq:1_2}) renders {\it intra} and {\it inter}-atomic terms. 
Among them, the one-center {\it intra}-atomic terms ($\alpha=\beta=\gamma$) have overwhelming weight,  followed by
the two-center terms, for which two of the atomic site indices are equal, that account for 2\% to 5\% of the SO splitting.\cite{Moores1973PRSL} 
The three-center terms ($\alpha\ne\beta\ne\gamma$)
are, thus,  disregarded altogether.
Due to the spherical symmetry, the {\it intra}-atomic matrix elements centered in an arbitrary atom $\alpha$ can be factored as
\begin{eqnarray} 
\langle is_i|H_{SO}^{intra}|js_j\rangle_\alpha =
\langle \mathcal{R}(r_\alpha)|\hat{\lambda}_\alpha(r_{\alpha})|\mathcal{R}(r_\alpha)\rangle
\langle l_im_is_i,\alpha|\dfrac{\mathbf{S}\cdot\mathbf{L}_\alpha}{\hbar^2}|l_jm_js_j,\alpha\rangle,
\label{eq:1_4}
\end{eqnarray}
where the radial part can be associated with the empirical $\lambda^{_{SOC}}_\alpha$ parameter
\begin{equation}
\langle \mathcal{R}(\mathbf{r}_\alpha)|\dfrac{\hbar^2}{2m^2c^2}\dfrac{1}{r_\alpha} 
\dfrac{dV_\alpha(r_\alpha)}{dr_\alpha} | \mathcal{R}(\mathbf{r}_\alpha)\rangle = 
\lambda^{_{SOC}}_\alpha.
\label{radialpart}
\end{equation}
We obtain the SOC constants $\lambda^{_{SOC}}_\alpha$ (Table S2) from the literature for the P, Sn and I elements.

Having defined the {\it intra}-atomic matrix elements, the {\it inter}-atomic ones can be obtained thereof (see Supporting Information), 
by using the Mulliken approximation for multicenter integrals.\cite{MullikenMultiCenter}
Therefore, we write the expression for the matrix elements of the spin-orbit coupling in the eH-TB framework as\cite{KirzcenowMn12}
\begin{eqnarray}
&\ &\langle is_i,\beta|H_{SO}|js_j,\gamma\rangle =
\delta_{\beta,\gamma} \sum_\alpha \sum_{k,l\in\alpha} S_{ik}(\beta,\alpha) S_{lj}(\alpha,\beta)
\langle ks_k|H_{SO}^{intra}|ls_l\rangle_\alpha + 
\label{HsoTB}\\
&\ &\left(1-\delta_{\beta,\gamma}\right) 
\left\{ 
\sum_{k\in\beta} S_{kj}(\beta,\gamma)\langle is_i|H_{SO}^{intra}|ks_k\rangle_\beta 
+
\sum_{k\in\gamma} S_{ik}(\beta,\gamma) \langle ks_k|H_{SO}^{intra}|js_j\rangle_\gamma
\nonumber
\right\},
\end{eqnarray}
The overlap matrix is block diagonal in the spinor Hilbert space. The SO interaction is considered on the same level as the $H_0$ hamiltonian.

Once we have the total hamiltonian $H=H_0+H_{SO}$, 
the generalized eigenvalue equation $H\boldsymbol{\Phi}=S\boldsymbol{\Phi}E_{diag}$ 
is solved to yield the molecular orbitals of the entire double-helix in the form of eigenpairs 
($\epsilon_n,\Phi_n$), written as spinors in the basis of the atomic STO's
\begin{equation}
\Phi_n(\mathbf{r}) = \sum_{\nu,s=\uparrow,\downarrow} A^n_{\nu s}\chi_{\nu s}(\mathbf{r}),
\label{spinor}
\end{equation}
where the combined index $\nu$ designates the pair of indices ($i,\alpha$).
Due to the spin degree of freedom, both $H_0$ and $S$ are block diagonal matrices, with each block associated with a 
different spin projection of the spinor. 
To calculate the electron propagation in the SnIP double-helix, we solve the time-dependent Schr\"odinger equation (TDSE) for an arbitrary inital 
quantum state $|\Psi\rangle$
\begin{equation}
i\hbar \dfrac{\partial}{\partial t} \left|\Psi(t) \right\rangle     
= \left(\hat{H}_0 + \hat{H}_{SO}\right) \left|\Psi(t) \right\rangle,
\label{TDSE}
\end{equation}
disregarding the nuclear motion. We write the arbitrary quantum state in terms of the energy eigenstates, 
as $\Psi(\mathbf{r}) = \sum_{n} C_n \Phi_n(\mathbf{r})$. Thus, the time evolution of $\Psi$ is given by
\begin{equation}
\Psi(\mathbf{r},t) = e^{-i\hat{H}t/\hbar}\Psi(\mathbf{r}) = \sum_n e^{-i\epsilon_n t /\hbar}C_n\Phi_n(\mathbf{r},0)~.
\label{Psi1}
\end{equation}
In order to get information about the spatial distribution of $\Psi(\mathbf{r},t)$ as well as its spin probability density
we use Eq. (\ref{spinor}), so that
\begin{equation}
\Psi(\mathbf{r},t) = \sum_n e^{-i\epsilon_n t /\hbar} C_n \sum_{\nu s} A^n_{\nu s} \chi_{i,\sigma}(\mathbf{r}) = 
\sum_{\nu s} Q_{\nu s}(t) \chi_{\nu s}(\mathbf{r})~,
\label{Psi(r,t)}
\end{equation}
where $Q_{\nu s}(t) = \sum_{n} \exp\left[{-i\epsilon_n t /\hbar}\right]C_n A^n_{\nu s}$, with $\nu$ designating the orbital quantum numbers and 
$s=\uparrow,\downarrow$ for the spin projection along the $z$ direction.
For convenience, we separate the wavefunction according to their spin components as
\begin{equation}
\Psi(\mathbf{r},t) = \sum_s \Psi_s(\mathbf{r},t) = 
\sum_\nu \left\{Q_{\nu\uparrow}(t) \chi_{\nu\uparrow}(\mathbf{r}) + Q_{\nu\downarrow}(t) \chi_{\nu\downarrow}(\mathbf{r}) \right\}.
\end{equation}

{\bf Spin probability current.}

It has been pointed out that the conventional spin current, usually defined as 
$ \mathbf{j}_s = \dfrac{\hbar}{m} \Im m \left[ \Psi^*_s \boldsymbol{\nabla} \Psi_s \right]$, or
alternatively as the mean value of the operator $(1/2)(\hat{\mathbf{v}}\hat{\mathbf{S}}+\hat{\mathbf{S}}\hat{\mathbf{v}})$,
is incomplete and unphysical under spin-flip hamiltonians.\cite{J-spinPRL,J-spinPRB,J-spinJapan,AJP-spin}
Thus, by considering the spin-orbit interaction, the continuity equation  for the probability density is written as
\begin{equation}
\begin{split}
\sum_{s=\uparrow,\downarrow}\frac{\partial |\Psi_s|^2}{\partial t} = 
&-\nabla \cdot\left\{\dfrac{\hbar}{m^*}\sum_{s=\uparrow,\downarrow} \Im m
\left[ \Psi^*_s \boldsymbol{\nabla} \Psi_s \right] \right\}\\
&- \nabla\cdot\left\{\dfrac{1}{2(m^*c)^2}\sum_{s,s'} 
\left[ \Psi^*_s \mathbf{S}_{s,s'} \Psi_{s'}
\right] \times \boldsymbol{\nabla} V(\mathbf{r})\right\}~,
\label{continuity}
\\
\end{split} 
\end{equation}
where we have omitted the time variable for the sake of clarity (details provided as Supporting Information). The LHS is simply the time-dependent total probability density whereas,  
on the RHS, we identify the conventional probability density current of well defined spin channel,
$
\mathbf{j}_s = 
\dfrac{\hbar}{m^*} \Im m
\left[ \Psi^*_s \boldsymbol{\nabla} \Psi_s \right] 
$,
and the spin-mixed probability density current that is due to the spin-orbit interaction,
\begin{equation}
\sum_{s'} \mathbf{j}^{_{SO}}_{s,s'} = 
\sum_{s'} \dfrac{1}{2(m^*c)^2} 
\left[ \Psi^*_s \mathbf{S}_{s,s'} \Psi_{s'}
\right] \times \boldsymbol{\nabla} V(\mathbf{r}).
\label{jso}
\end{equation}
The later has also been associated with the torque dipole density\cite{J-spinPRL} or the angular spin current density.\cite{J-spinPRB}
The effective mass $m^*$ is a consequence of the tight-binding formalism. 
Equation (\ref{continuity}) essentially  connects the quantum dynamics, on the LHS, with the measurable current densities on the RHS.
It can be written in terms of the probability current densities as 
\begin{equation}
\sum_{s=\uparrow,\downarrow}\frac{\partial \rho_ s}{\partial t} = 
-  \boldsymbol{\nabla}\cdot \sum_{s}    \left\{ \mathbf{j}_s 
+ \sum_{s'} \mathbf{j}^{_{SO}}_{s,s'} \right\}~.
\end{equation}

To calculate the electronic transport in the double helix we integrate the continuity equation over a volume $\mathcal{V}$ along the nanowire axis 
(Figure S1) and apply the divergence theorem to obtain
\begin{equation}
\frac{\partial}{\partial t}  \left[ P_\uparrow + P_\downarrow \right]_\mathcal{_{V}} = 
-\sum_{s} \int_\mathcal{_{S}} {j}_{z,s} dA_{\perp}
-  \int_\mathcal{_{S}} 
\left[\sum_{s,s'} j^{_{SO}}_{s,s'}\right]_z dA_{\perp}~,
\label{detector}
\end{equation}
where $P_s$ is the spin-dependent electron population inside $\mathcal{V}$, associated with the spinor wavepackets $\Psi_s(\mathbf{r},t)$.
On the RHS, ${j}_{z,s} dA_{\perp}$ is the probability density flux with well defined spin projection across the surface caps perpendicular do the $\hat{z}$
direction. The last term accounts for the probability density flux of the mixed-spin current.
At this point, we can write down an expression for the mixed-spin current 
(see Supporting Information). In order to have consistency between the TB matrix element  of Eq. (\ref{eq:1_4}) and 
the spin-orbit hamiltonian of Eq. (\ref{Hso}) we assume that the SOC constant is given by a Coulomb potential that is generated by an effective charge 
$Q^\prime_\alpha$, for each of the atomic species  in the double-helix.
That is
\begin{equation}
\boldsymbol{\nabla} V(\mathbf{r}) \approx \sum_\alpha \boldsymbol{\nabla} V_\alpha(\mathbf{r}-\mathbf{R}_\alpha)
\approx \sum_\alpha 
\dfrac{-e}{r_\alpha} 
\dfrac{d}{dr_\alpha}
\left(\dfrac{Q^\prime_\alpha}{r_\alpha}\right)\mathbf{r}_\alpha~.
\label{Qeff}
\end{equation}
Then, we substitute Eq. (\ref{Qeff}) into Eq. (\ref{radialpart}),
which is  integrated  over the radial part of the STO's to yield the effective charges in terms of the SOC constants (Table S2) that are obtained from the 
literature.  As a result, we obtain a spin-orbit interaction operator that is consistent with the eH-TB formalism.
Thus, the mixed-spin current flux along the axis of the double-helix can be explicitly written 
as
\begin{equation}
\left[\sum_{s,s'} j^{_{SO}}_{s,s'}\right]_z
\approx \dfrac{e\hbar}{2(m^*c)^2}
\left(
\Re e(\Psi^*_\uparrow \Psi_\downarrow) \sum_\alpha 
\dfrac{Q^\prime_\alpha y_\alpha}{r_\alpha^3}
-\Im m (\Psi^*_\uparrow \Psi_\downarrow) \sum_\alpha 
\dfrac{Q^\prime_\alpha x_\alpha}{r_\alpha^3}
\right)~,
\label{fluxz}
\end{equation}
with $\Psi_s(\mathbf{r},t)$ given by Eq. (\ref{Psi(r,t)}). 

\begin{figure}[htbp]
    \centering
 \includegraphics[width=0.85\linewidth]{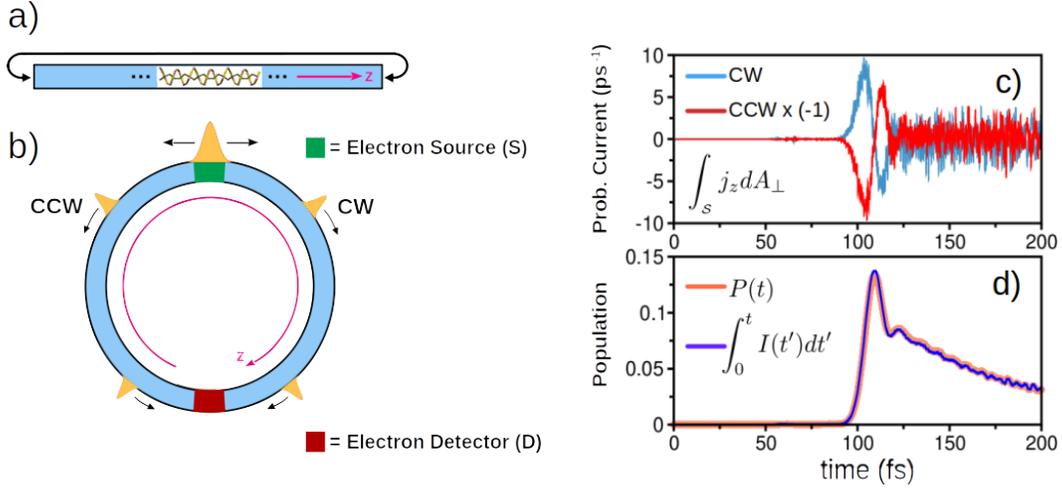}                             
 \caption{a) SnIP double-helix oriented along the $\hat{z}$ direction with periodic boundary conditions. b) The SnIP periodic structure is topologically equivalent 
to a ring. The green segment designates the electron source (S), where $\Psi(\mathbf{r},t=0)$ is created (or injected). The red segment designates the drain, where the 
electron in detected (D). The initial wavepacket (orange) splits in two wavepackets that propagate in opposite directions, towards the detector.
c) Wavepacket dynamics in an M (left-handed) SnIP double-helix without spin-orbit interaction.
Probability current for the CW (blue) and CCW (red) spinless wavepackets; the later multiplied by the factor $-$1 to improve clarity.
b) Time-dependent electronic population in the detector, without the $j^{_{SO}}_{s,s'}$ term.
}
\label{ring}
\end{figure}

{\bf Spin dependent electron transport.}
Let us consider a long strand of the SnIP double-helix, either of M (left-handed) or P (right-handed) chirality, 
subject to periodic boundary conditions, so that it is topologically 
equivalent to a 
double-helix ring, without deforming its original geometric structure. Figure \ref{ring} illustrates the concept.
Herein, for purposes of calculations, we assume an M-SnIP double-helix composed of 97 unit cells.
We consider that the initial wavepacket $\Psi(\mathbf{r},t=0)$ is created in a small portion of the strand, denominated source (S), 
which is comprised of 3 SnIP unit cells, as depicted in green in Figure \ref{ring}-b. 
In practice the source segment could be associated with an electrode, or a scanning tunneling microscope (STM) tip that injects electrons in the double-helix.
Since the initial wavepacket $\Psi(\mathbf{r},t=0)$ is not a stationary state of the entire strand, 
by solving the TDSE we observe that it splits in two symmetrical wavepackets that propagate in opposite directions. 
After a "time-of-flight" both travelling 
wavepackets are detected in the detector segment (D), which is equally distant from S by clockwise (CW) or counter-clockwise (CCW) paths. 
The detector segment is comprised of 4 SnIP unit cells, as depicted in red in Figure \ref{ring}-b. 
The detection of the travelling wavepackets in D is calculated with the continuity equation,
Eq. (\ref{detector}). From the experimental point of view, the electron detection could be accomplished by electric contacts or luminescent probes.
The same calculations were also performed on P-SnIP double-helices (see Supporting Information).

Before we consider the spin-dependent transport, it is instructive to start by  considering the  spinless situation, in which case $H_{SO}$ is disregarded.
The results are presented in Figure \ref{ring}, where panel c) shows the probability current for the CW (blue) and CCW (red) wavepackets as a function of time
and  panel d) shows the time-dependent electronic population in the detector, as given by the LHS and the RHS of Eq. (\ref{detector}), without the 
$\mathbf{j}^{_{SO}}_{s,s'}$ term. 
These results can be summarized as follows: 
1) there is no observable difference for spinless wavepackets travelling along CW or CCW directions on M (Figure \ref{ring}-c) or P-SnIP (Figure S2) 
strands; 
2) the continuity equation
derived for the TB formalism is consistent with the quantum dynamics calculations (Figure \ref{ring}-d); 
3) the time-of-flight the wavepackets take to travel 45 unit cells ($\approx$ 35.7 nm) in the absence of SOC is
$\tau_{tof}$ = 109 fs; and  
4) by equating the LHS with RHS of Eq. (\ref{detector}) we obtain a tight-binding effective mass of $m^* \approx 0.42~m_e$, which is 
similar to the value of 0.28 $m_e$ that was obtained from
ab-initio band structure calculations.\cite{purschke2021}
Thus, ignoring the SOC, the results for M-SnIP and P-SnIP strands show no observable difference (Figure S2).

In the remainder, we look into the effects caused by the interplay of the spin-orbit coupling with the chirality of the strands. 
Let us consider, for the sake of the argument, an initial state $\Psi(0)=\Psi_{s}$  in the source segment, with pure spin state $s_z = +1$  
parallel to the $\hat{z}$ axis of the M-SnIP double-helix. 
For practical reasons, we define the usage: \{M($-$),P($+$)\} for designating the SnIP axial chirality, \{CW,CCW\} for the wave propagation direction 
and $s$=\{$\uparrow,\downarrow$\} for spin orientation parallel or anti-parallel to the $\hat{z}$ axis.
The spin-helicity of the propagating wavepacket is given by 
$h = \mathbf{S}\cdot\mathbf{v}/(|\mathbf{S}||\mathbf{v}|) \equiv \pm 1$, where $\mathbf{S}$ is the spin vector and $\mathbf{v}$ is the propagation velocity of the 
wavepacket, independently of the chirality of the strand. 
According to this definition, the CW wavepacket has {\bf initially} $h^{\uparrow}_{_{CW}} = +1$ whereas 
the CCW one has $h^{\uparrow}_{_{CCW}} = -1$, so that $\hat{\mathcal{P}}\hat{\mathcal{T}}h^{\uparrow}_{_{CW}}=h^{\downarrow}_{_{CW}}=-h^{\uparrow}_{_{CW}}$.
Applying the time-reversal operation on the initial wavepackets we get the Kramers doublets:
$\{h^{\uparrow}_{_{CW}},h^{\downarrow}_{_{CCW}}\} = +1$ and 
$\{h^{\uparrow}_{_{CCW}},h^{\downarrow}_{_{CW}}\} = -1$.
In a free (achiral) medium, the dynamics of the wavepackets belonging to either of the doublets is equal, such as in the case of a beam
of circularly polarized light propagating in an achiral medium.
However, we ought to incorporate the chirality of the medium in the description.
To do so, we define the parity-even symmetry index $\eta = Ch \equiv \pm 1$, where C stands for \{M($-$),P($+$)\}. It combines 
the spin-helicity of the electron with  the chirality of the medium, rendering for the initial wavepackets  $\eta^{M\uparrow}_{_{CW}} = -1$ and 
$\eta^{M\uparrow}_{_{CCW}} =+1$.
By classifying the $\eta$ index  with respect to the $\mathcal{P}\mathcal{T}$ symmetries we get  
$\hat{\mathcal{P}}\hat{\mathcal{T}}\eta^{M\uparrow}_{_{CW}}=\eta^{P\downarrow}_{_{CW}} =-1$, and
$\hat{\mathcal{P}}\hat{\mathcal{T}}\eta^{M\uparrow}_{_{CCW}}=\eta^{P\downarrow}_{_{CCW}}=+1$, and so on.
Thus, considering all possible scenarios, {\it i.e.} the forward and backward propagating states of both spin orientations on helical strands of M and P 
chirality, the Kramers doublet is augmented to give rise to a fourfold doublet with 
$\{\eta^{M\uparrow}_{_{CW}},\eta^{M\downarrow}_{_{CCW}},\eta^{P\downarrow}_{_{CW}},\eta^{P\uparrow}_{_{CCW}}\} \equiv \eta^-$
and
$\{\eta^{M\uparrow}_{_{CCW}},\eta^{M\downarrow}_{_{CW}},\eta^{P\downarrow}_{_{CCW}},\eta^{P\uparrow}_{_{CW}}\} \equiv \eta^+$.

\begin{figure}[htbp]
    \centering
 \includegraphics[width=0.85\linewidth]{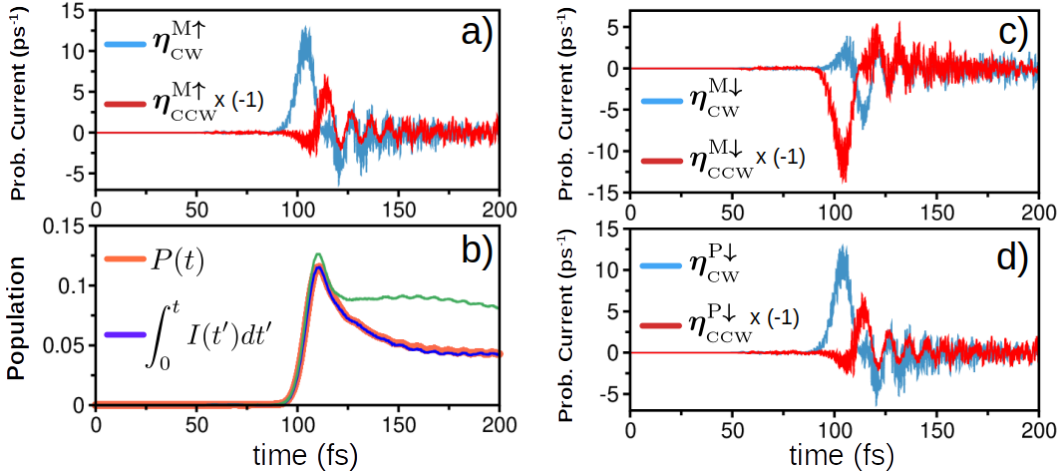}                             
 \caption{
a) Probability density currents at the detector segment D produced by the counter-propagating wavepackets revealing the asymmetry of propagation velocity.
Probability current for the $\eta^{M\uparrow}_{_{CW}} =\eta^-$ (blue) and $\eta^{M\uparrow}_{_{CCW}} =\eta^+$ (red) wavepackets, the later multiplied by the $-1$ 
factor for the sake of clarity. Note that $v(\eta^{M\uparrow}_{_{CW}}) > v(\eta^{M\uparrow}_{_{CCW}})$.
b) Time-dependent electronic population in the detector, as given by the LHS (orange) and the RHS (blue) terms  of Eq. (\ref{detector}).
The green curve, obtained without the $\mathbf{j}^{_{SO}}_{s,s'}$ term, evinces the relevance of the spin-mixed component of $\mathbf{j}$.
Results for different counter-propagating wavepackets:
c) $v(\eta^{M\downarrow}_{_{CCW}}) > v(\eta^{M\downarrow}_{_{CW}})$ and d) $v(\eta^{P\downarrow}_{_{CW}}) > v(\eta^{P\downarrow}_{_{CCW}})$.
Note that $v(\eta^{M\downarrow}_{_{CCW}}) = v(\eta^{P\downarrow}_{_{CW}})$.}
\label{SOC}
\end{figure}

After these general considerations, we now analyse the transport properties of electrons in a chiral SnIP strand taking into account the SO interaction. 
Figure \ref{SOC}-a shows that the probability current due to the wavepacket $\eta^{M\uparrow}_{_{{\bf CW}}}=-1$ is faster and has higher fluence rate than that of the 
counter-propagating wavepacket with $\eta^{M\uparrow}_{_{{\bf CCW}}}=1$.
The same behavior is evinced for the probability current of the $\eta^{M\downarrow}_{_{{\bf CW,CCW}}}$ pair of 
contra-propagating wavepackets (Figure \ref{SOC}-c)
and also for the $\eta^{P\uparrow}_{_{{\bf CW,CCW}}}$ pair (Figure \ref{SOC}-d).  
The consistency of the asymmetric propagation effect with the continuity equation is evinced in Figure \ref{SOC}-b, by the agreement between 
the time-dependent population (orange curve) with the probability density flux integrated over time (blue curve).
It is also interesting to realize the relevance of $\mathbf{j}^{_{SO}}_{s,s'}$ for the continuity equation, as revealed by the green curve in Figure \ref{SOC}-b
that describes the time-integrated probability density flux without the spin-mixed term.

However, spin is not a paramount cause of the effect. 
In fact, a propagation velocity asymmetry has also been reported for electrons in SWCNT, neglecting altogether the spin-orbit interaction, 
but considering instead the  orbital angular momentum of the carriers associated with the $K$ and $K^\prime$ valleys. 
In this case the effect is produced by a curvature-induced effective spin-orbit interaction, and it is revealed after second order perturbation treatment.
In a theoretical study Izumida et al.\cite{SWCNT-Izumida} showed that left- and right-propagating electrons in the same K 
(or K$^\prime$) valley have different velocities in 
chiral and armchair nanotubes.  It was shown that $v_L^{(K)} > v_R^{(K)}$ but  $v_R^{(K)} = v_L^{(K^\prime)}$, which can be mapped onto our 
results if $K(K^\prime)\iff \uparrow(\downarrow)$ and $L(R)\iff CW(CCW)$, so that 
$v(\eta^{M\uparrow}_{_{CW}}) > v(\eta^{M\uparrow}_{_{CCW}})$ but $v(\eta^{M\uparrow}_{_{CW}}) = v(\eta^{M\downarrow}_{_{CCW}})$.
In general we have $v(\eta^-) > v(\eta^+)$ for the elements of the previously defined fourfold Kramers doublet.
Yet another analogous phenomenon is the so called planar chirality effect,\cite{AsymmetricV-PRL} whereby the transmission of circularly polarized EM waves through 
a planar chiral structure undergoes asymmetric propagation depending on the direction of wave propagation; 
note that the handedness of a planar chiral structure is reversed when observed from opposite sides of the structure's plane.

\begin{figure}[htbp]
    \centering
 \includegraphics[width=0.85\linewidth]{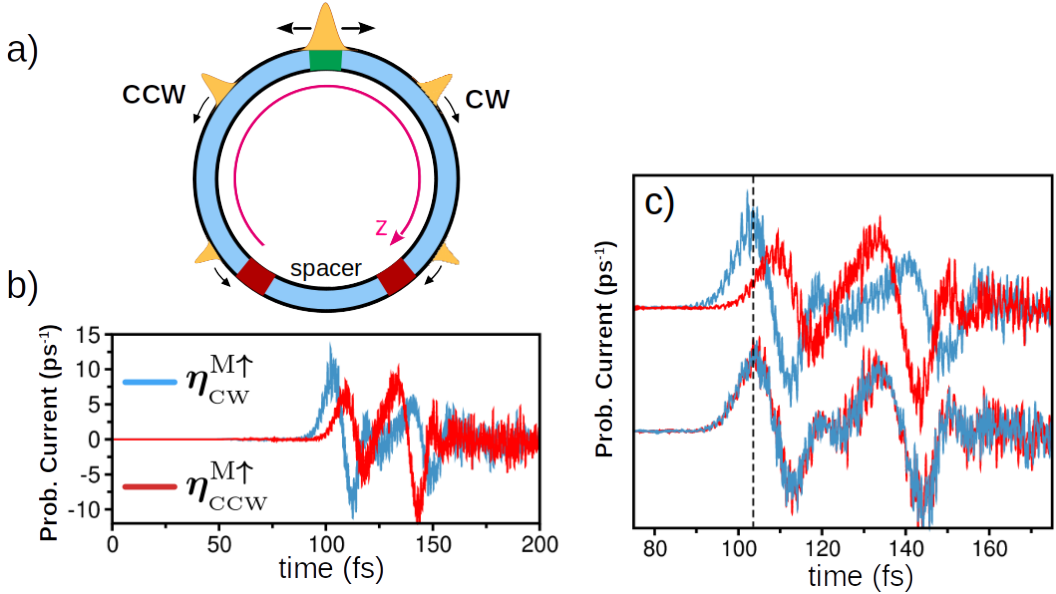}                             
 \caption{a) Periodic M-SnIP double-helix. 
Two detector segments are considered, designated D$_\text{CCW}$ and 
D$_\text{CW}$, separated by a spacer of length equal to 79.34 \AA.
b) Probability current for the $\eta^{M\uparrow}_{_{CW}} = -1$ (blue) and $\eta^{M\uparrow}_{_{CCW}} =+1$ (red) wavepackets.
The time-of-flight is $\tau_{tof}(\eta^{M\uparrow}_{_{CW}}) \approx$ 103 fs and $\tau_{tof}(\eta^{M\uparrow}_{_{CCW}}) \approx$ 109 fs.
c) Comparison between the spin-dependent (shifted up) and the  spinless (shitfed down) propagation dynamics, evincing that 
$v(\eta^-) \approx v_0$, where $v_0$ corresponds to the spinless case.
}
\label{spacer}
\end{figure}

The ring structure of Figure \ref{ring} is not well suited to determine the propagation velocities of the counter-propagating wavepackets. 
Thus, we modified 
its structure to accommodate two detector segments, D$_\text{CCW}$ and D$_\text{CW}$, both with the same size as the original D, 
but separated by a spacer segment of length 79.34 \AA~(shown in Figure \ref{spacer}-a). 
The distance between the source (S) and the detectors (D$_\text{CW}$ and D$_\text{CCW}$) is preserved (35.7 nm) as in the original 
structure. Then, we apply the continuity equation, Eq. (\ref{detector}), to both detectors to evaluate the probability flux due to the counter propagating wavepackets
(Figure \ref{spacer}-b).
The positive peaks indicate the inflow of charge in the detector whereas the negative peaks are associated with the outflow.
The time-of-flight for each case is $\tau_{tof} \approx$ 103 fs for the $\eta^{M\uparrow}_{_{CW}}$ wavepacket and $\tau_{tof} \approx$ 109 fs for the 
$\eta^{M\uparrow}_{_{CCW}}$ wavepacket, with the later being approximately 6\% slower.
Panel c) of Figure \ref{spacer} compares the spin-dependent to the  spinless propagation dynamics for this setup. 
We notice that $v(\eta^-) \approx v_0$, where $v_0$ corresponds to the spinless case, whereas $v(\eta^+)$ decreases,
so that we have $v(\eta^-) \approx v_0 > v(\eta^+)$. Additional simulations performed for different ring setups (Figure S7) and different elements of the
Kramers doublets corroborate the effect.

{\bf Discussions.}
We have shown  that the mobility of electrons in structures with axial chirality depends on 3 factors, as described by the $\eta$ index:
the direction of the spin angular momentum (or for that matter an axial vector), the direction of motion and the chirality of the 
underlying structure, the three of them combined through a spin-orbit like interaction. As a result, we have $v(\eta^-) > v(\eta^+)$, as previously described,
in qualitative agreement with the CISS phenomena.\cite{SecondCISS}
Due to its generality, this effect should occur in a variety of systems that satisfy the aforementioned basic conditions.
Scattering and decoherence perturbations, though, not taken into consideration in the present unitary quantum description,
should also interfere with the measured effect.
The environmental coupling (including defects and the vibrational normal modes of the structure) shall decrease considerably the coherence length of the electron. 
However, the spin polarization lifetime exceeds the transport time by several orders of magnitude.
Thus, the spin-dependent quantum mobilities of the left- and right-propagating electrons in the chiral structure provide the basis for 
employing semiclassical frameworks, like the kinetic Monte Carlo method, to calculate the CISS effect in general situations.
Another issue, not yet resolved, concerns the importance of the strength of the SO interaction. In SnIP, the elements Sn and I have strong intrinsic SOC constants,
unlike the organic elements. However, for organic systems, the effective SO hamiltonian may be augmented by external electric fields due, for 
instance, to hydrogen bonds\cite{Varela} or the influence of strain in the SO coupling.\cite{GuineaPRB,Guinea-NanoPhys} 
Yet another possibility is that the CISS effect has a nonlinear dependence on the 
SOC strength, evidenced perhaps in tunnel/hopping models.\cite{VarelaPRB}

Finally, once separation and stabilization of the SnIP enantiomers is attained, this material could also be envisaged as a gyrotropic metamaterial in the form of 
thin films or coating, for it has recently been demonstrated that 3D-chiral metamaterials with strong chirality can exhibit unconventional optical properties 
such as negative refractive index 
and enhanced optical activity, among others.\cite{AsymmetricV-PRL,MetamaterialsPRB,MetamaterialsPRL}

{\bf Conclusions.}
In summary, based on general quantum mechanical principles we have derived a spin-current continuity equation using the tight-binding formalism, and 
showed its consistency with quantum dynamics calculations. The formalism was applied to the promising double-helix system SnIP. We derived an expression for the 
spin-mixed current and showed its importance for the spin conservation in spin transport experiments. Finally we have demonstrated the chirality induced 
propagation velocity asymmetry effect 
for charge transport in chiral systems and analysed its relevance to the chiral-induced spin selectivity (CISS) and other phenomena.

{\bf Acknowledgements.}
This study was financed by Coordena\c c\~ao de Aperfei\c coamento de Pessoal de N\'{\i}vel Superior Brasil (CAPES) - Finance Code 001, 
by the Brazilian National Counsel of Technological and Scientific Development (CNPq) and the National Institute for Organic Electronics (INEO).
L.G.C.R. acknowledges allocation of supercomputer time from Laboratory for Scientiﬁc Computing (LNCC/MCTI, Brazil).

\bibliography{myRefs}
\bibliographystyle{apsrev4-1}

\pagebreak
\begin{center}
\textbf{\large Supporting Information for \\
Chirality Induced Propagation Velocity Asymmetry}
\end{center}

\setcounter{equation}{0}
\setcounter{figure}{0}
\setcounter{table}{0}
\setcounter{page}{1}
\makeatletter
\renewcommand{\theequation}{S\arabic{equation}}
\renewcommand{\thefigure}{S\arabic{figure}}
\renewcommand{\bibnumfmt}[1]{[S#1]}
\renewcommand{\citenumfont}[1]{S#1}

\section{Tight-Binding Hamiltonian}

We compute the tight-binding (TB) hamiltonian in the basis of atomic Slater-type orbitals (STO's), 
$|\chi_{nlm}s\rangle$, 
where $i=\{nlm\}$ represents the orbital quantum numbers associated with  the wavefunction $\chi_{nlm}(\mathbf{r}_\alpha)=\mathcal{R}_n(r_\alpha) Y^l_m(\theta,\varphi)$,
and $s=\{-1,1\}$ (lowercase $s$) designates the spin state of the electron along the $z$ direction. 
In particular, the wavefunction components $\mathcal{R}_n(r_\alpha)$ and 
$Y_{lm}(\theta,\varphi)$ describe, respectively, the radial and angular parts of the STO for an atom located at position $\mathbf{R}_\alpha$, with 
$\mathbf{r}_\alpha = \mathbf{r} - \mathbf{R}_\alpha$, 
\begin{equation} 
\langle \mathbf{r}_\alpha | n \rangle = \sqrt{\frac{2 \zeta}{(2n)!}} (2 \zeta)^{n}~
r_\alpha^{n-1}~e^{-\zeta r_\alpha}~Y_{lm} (\theta, \varphi) . 
\end{equation} 
The basis set includes the 3s and 3p atomic orbitals for the phosphorus (P) atoms and the 5s and 5p orbitals for the tin (Sn) and iodine (I) atoms.
Several studies have used the TB method, either in the Slater-Koster\cite{GuineaPRB,Medina2016,Hedegard2019,Cuniberti} (SK)
or the extended-H\"uckel\cite{Cerda,KirzcenowMn12,PhysRevBRenani2013} (eH)
framework, to describe carbon nanotubes, DNA molecules and nanosystems alike.
We adopt the extended-H\"uckel approach. 

The total hamiltonian is comprised of a spinless part and the spin-orbit interaction part.
The matrix elements of the spinless semiempirical extended-H\"uckel hamiltonian are defined as 
$H_{ij} = \kappa_{ij}S_{ij}$, where $\kappa_{ij} = K_{ij}(\varepsilon_i+\varepsilon_j)/2$
and $S_{ij}(\alpha,\beta) = \langle\chi_i(\mathbf{r}_\alpha)|\chi_j(\mathbf{r}_\beta)\rangle$ is the overlap between
atomic orbitals $i$ and $j$ located at $\mathbf{R}_\alpha$ and $\mathbf{R}_\beta$, respectively, with
$S_{ij}=\delta_{ij}$ when $\alpha=\beta$.
The empirical parameters $\varepsilon_i$ are generally associated with the negative ionization energies of the valence atomic orbitals and the 
parameter $K_{ij}$ is the Wolfsberg–Helmholz empirical constant. 
We implement the modified Wolfsberg-Helmholz formula,\cite{COM} 
$K_{ij} = \overline{k}_{ij} + \Delta^2 + \Delta^4(1-\overline{k}_{ij})$, with $\overline{k}_{ij}=(k_{_{\text{WH}},i}+k_{_{\text{WH}},j})/2$  and 
$\Delta = \left(\varepsilon_i-\varepsilon_j\right)/\left(\varepsilon_i+\varepsilon_j\right)$. 
The STO parameters used in this work are shown in Table \ref{EH_optparms1}. 

\begin{table}[p]
\centering
\caption{Extended H\"uckel parameters used for the SnIP double-helix. }
\label{EH_optparms1}
		\begin{tabular}{c|c|c|c|c|c|c} 
\hline
\hline
\textbf{EHT-Symbol}& $\mathbf{N_{val}}$ & \textbf{n} & \textbf{spdf} & $\varepsilon_{i}$ (eV) & $\mathbf{\zeta}$($a_0^{-1}$) & $\mathbf{k_{WH}}$ \\
\hline
P    & 5 & 3 & s & -18.077 & 2.028 & 1.170 \\ 
P    & 5 & 3 & p & -14.016 & 2.160 & 2.872 \\ 
\hline
Sn   & 4 & 5 & s & -14.975 & 2.118 & 1.809 \\
Sn   & 4 & 5 & p & -8.186  & 2.159 & 2.452 \\ 
\hline
I    & 7 & 5 & s & -17.955 & 3.038 & 0.150 \\ 
I    & 7 & 5 & p & -13.190 & 2.426 & 2.571 \\ 
\hline
\hline
\end{tabular}
\end{table}

The total hamiltonian can be divided into a spinless part ($H_0$), associated with the eH-TB hamiltonian,                                               
and a spin-orbit (SO) interaction part
\begin{equation}
H_{SO} = \frac{1}{2(mc)^2} \mathbf{S}\cdot\boldsymbol{\nabla} V(\mathbf{r})\times \mathbf{p},
\label{Hso}
\end{equation}
where $\mathbf{S}=\tfrac{\hbar}{2}\boldsymbol{\sigma}=\tfrac{\hbar}{2}(\sigma_x,\sigma_y,\sigma_z)$ is the spin vector operator comprised of the $\sigma$ 
Pauli matrices, 
$\mathbf{p}=-i\hbar\boldsymbol{\nabla}$ is the electron momemtum operator and $V(\mathbf{r})$ is the total
Coulomb potential energy felt by the electron.
We follow the approach of  Renani and Kirzcenow\cite{KirzcenowMn12,PhysRevBRenani2013}
to write the spin-orbit hamiltonian in the framework of the eH-TB formalism.
Despite  the complexity of the SO operator, the main contribution of the SO interaction arises from the one-electron operators                                          
that couple the electron with the nuclei.\cite{Moores1973PRSL}
Therefore, in the TB method, the total Coulomb potential can be approximated by a summation over spherically 
symmetric single particle potentials produced by all the nuclei $\alpha$, 
$V(\mathbf{r}) \approx \sum_\alpha V_\alpha(|\mathbf{r}-\mathbf{R}_\alpha|)$,\cite{SlaterKoster-TB} so that 
\begin{equation}
H_{SO} \approx \sum_\alpha \dfrac{1}{2m^2c^2}\dfrac{1}{|\mathbf{r}_\alpha|} 
\dfrac{dV_\alpha(r_\alpha)}{dr_\alpha} \mathbf{S}\cdot\mathbf{L}_\alpha
= \sum_\alpha \hat{\lambda}_\alpha(r_\alpha) \frac{\mathbf{S}\cdot\mathbf{L}_\alpha}{\hbar^2},
\label{SO1}
\end{equation}
with $\mathbf{L}_\alpha = (\mathbf{r} -\mathbf{R}_\alpha)\times \mathbf{p}$ describing the angular momemtum of the electron with respect to atom $\alpha$.
It has been shown that the effective single particle SO operator $\xi_\alpha \mathbf{S}\cdot\mathbf{L}_\alpha$ produces very good results\cite{Moores1973PRSL} 
when the coupling parameter $\xi_\alpha$ is obtained from ab-initio calculations or experimental fits, so as to incorporate SO many-body effects.

In the following, we make use of the transformation relations between the cubic harmonics (default in TB representation) and the spherical harmonics.
The matrix elements of  $H_{SO}$ can be written as 
\begin{equation}
\langle is_i,\beta|H_{SO}|js_j,\gamma\rangle=\sum_{\alpha}\langle is_i,\beta|\left(\hat{\lambda}_\alpha(r_{\alpha})
\dfrac{\mathbf{S}\cdot\mathbf{L}_\alpha}{\hbar^2}\right)|js_j,\gamma\rangle,
\label{eq:1_2}
\end{equation}
with $\beta$ and $\gamma$ designating the atoms located at positions $\mathbf{R}_\beta$ and $\mathbf{R}_\gamma$, which have orbital quantum numbers 
$i\equiv\{n_il_im_i\}$ and $j\equiv\{n_jl_jm_j\}$, respectively, as well as spin states $s_i$ and $s_j$. 
The matrix element of Eq. (\ref{eq:1_2}) renders {\it intra} and {\it inter}-atomic terms. 
Among them, the one-center {\it intra}-atomic terms ($\alpha=\beta=\gamma$) have overwhelming weight,  followed by
the two-center terms, for which two of the atomic site indices are equal, that account for 2\% to 5\% of the SO splitting.\cite{Moores1973PRSL} 
The three-center terms ($\alpha\ne\beta\ne\gamma$)
are, thus,  disregarded altogether.
Therefore, we have
\begin{eqnarray}
\langle is_i,\beta|H_{SO}|js_j,\gamma\rangle&=&
\delta_{\beta,\gamma}\sum_\alpha\langle is_i,\beta|\left(\hat{\lambda}_{\alpha}(r_{\alpha})\dfrac{\mathbf{S}\cdot\mathbf{L}_\alpha}{\hbar^2}\right)|js_j,\beta\rangle
\label{eq:1_3}\\
&+&
(1-\delta_{\beta,\gamma})
\bigg[\langle is_i,\beta|\left(\hat{\lambda}_{\beta}(r_{\beta})\dfrac{\mathbf{S}\cdot\mathbf{L}_\beta}{\hbar^2}\right)|js_j,\gamma\rangle
+
\langle is_i,\beta|\left(\hat{\lambda}_{\gamma}(r_{\gamma})\dfrac{\mathbf{S}\cdot\mathbf{L}_\gamma}{\hbar^2}\right)|js_j,\gamma\rangle\bigg].
\nonumber
\end{eqnarray}

We start by writting down the {\it intra}-atomic matrix element centered in atom $\alpha$.
Due to the spherical symmetry, the {\it intra}-atomic matrix elements can be factored as
\begin{eqnarray} 
\langle is_i|H_{SO}^{intra}|js_j\rangle_\alpha &=&
		\langle is_i,\alpha|\hat{\lambda}_\alpha(r_{\alpha})\dfrac{\mathbf{S}\cdot\mathbf{L}_\alpha}{\hbar^2}|js_j,\alpha\rangle\nonumber\\
&=&\langle \mathcal{R}(r_\alpha)|\hat{\lambda}_\alpha(r_{\alpha})|\mathcal{R}(r_\alpha)\rangle
\langle l_im_is_i,\alpha|\dfrac{\mathbf{S}\cdot\mathbf{L}_\alpha}{\hbar^2}|l_jm_js_j,\alpha\rangle.
\label{eq:1_4}
\end{eqnarray}
The angular part of Eq. (\ref{eq:1_4}) preserves $L_\alpha$ and $S$, so that it is non-zero only for $l_i=l_j$,\cite{condon1935theory} 
\begin{eqnarray}
\langle l_im_is_i,\alpha|\mathbf{S}\cdot\mathbf{L}_\alpha|l_jm_js_j,\alpha\rangle &=&
\dfrac{\hbar^2}{2}~s_i m_i~ \delta^{l_i,m_i,s_i}_{l_j,m_j,s_j}
\label{angular}\\
&+&\dfrac{\hbar^2}{2}
\sqrt{l_j(l_j+1)-m_j(m_j+s_j)}(1-\delta_{s_i,s_j})\delta^{l_i,m_i}_{l_j,m_j+s_j}~.
\nonumber
\end{eqnarray}
The radial part of Eq. (\ref{eq:1_4}) can be associated with the empirical $\lambda_{SOC}$  parameter
\begin{equation}
\langle \mathcal{R}(r_\alpha)|\hat{\lambda}_\alpha(r_{\alpha})|\mathcal{R}(r_\alpha)\rangle = \lambda^{_{SOC}}_\alpha~.
\label{radial}
\end{equation}
We obtain the parameters $\lambda^{_{SOC}}_\alpha$ from the literature, as shown in Table \ref{SOCC}, for the P, Sn and I elements.
Hund's rules are adopted.\cite{condon1935theory}
\begin{table*}
\caption{Intra-atomic spin-orbit coupling paramter. To obtain $\lambda_{SOC}$ for P an effective principal quantum number $n_{eff}$ = 1.51 was used 
\cite{Neff-P-CUSACHS} in conjunction with the parameters  of \citet{SOC-P}.} 
\begin{tabular}{c|c|c|c}
\hline \hline
 Element                 &
 $\lambda_{SOC}$ (eV)    &
 $\lambda_{SOC}$ (cm$^{-1}$)    &
 Ref.                    \\ 
\hline
P   & -0.00889  & -71.7   & \citet{SOC-P}  \\ 
Sn  & 0.68     & 5484.6      & \citet{SOC-Sn-PhysRevB}  \\
I   & -0.62848 & -5069.0  & \citet{SOC-I} \\
\hline
\hline
\end{tabular}
\label{SOCC} 
\end{table*}

Therefore, by combining the results of Eq. (\ref{angular}) and Eq. (\ref{radial}), we get for the {\it intra}-atomic matrix elements
\begin{equation}
\langle is_i|H_{SO}^{intra}|js_j\rangle_\alpha =
\dfrac{\lambda^{_{SOC}}_\alpha}{2}\left 
\{m_i s_i~ \delta_{i,j} + 
\sqrt{l_j(l_j+1)-m_j(m_j+s_j)}(1-\delta_{s_i,s_j})\delta^{l_i,m_i}_{l_j,m_j+s_j} \right\}.
\label{intra}
\end{equation}

For the {\it inter}-atomic matrix elements of Eq. (\ref{eq:1_3}) we have
\begin{equation}
\langle is_i,\beta|H_{SO}^{inter}|js_j,\gamma\rangle=
\langle is_i,\beta|\hat{\lambda}_{\beta}(r_{\beta})\dfrac{\mathbf{S}\cdot\mathbf{L}_\beta}{\hbar^2}|js_j,\gamma\rangle
+\langle is_i,\beta|\hat{\lambda}_{\gamma}(r_{\gamma})\dfrac{\mathbf{S}\cdot\mathbf{L}_\gamma}{\hbar^2}|js_j,\gamma\rangle,
\label{inter}
\end{equation}
with $\beta\neq\gamma$. Since the one-center terms are much larger than the two-center ones,\cite{Moores1973PRSL} we can
use the Mulliken-R\"udenberg approximation\cite{MullikenMultiCenter,Rudenberg}   
to simplify the two-center integrals in terms of one-center integrals.
In addition to the size difference, 
the fact that the radial part of the SO interaction decays approximately as $r_\alpha^{-3}$ indicates that the SO interaction is localized
in space. Now, to be consistent with the eH formalism and the valence-electron asumption, the summation is carried out over the orthonormal basis set 
$\{k\}$ comprised of valence orbitals associated with the atoms $\beta$ and $\gamma$. Thus, we obtain for Eq. (\ref{inter})
\begin{eqnarray}
\langle is_i,\beta|H_{SO}^{inter}|js_j,\gamma\rangle &=&
\sum_{k\in\beta}\langle is_i,\beta|\hat{\lambda}_{\beta}(r_{\beta})\dfrac{\mathbf{S}\cdot\mathbf{L}_\beta}{\hbar^2}|ks_k,\beta\rangle
\langle ks_k,\beta|js_j,\gamma\rangle \nonumber\\
&+&\sum_{k\in\gamma}\langle is_i,\beta|ks_k,\gamma\rangle
\langle ks_k,\gamma|\hat{\lambda}_{\gamma}(r_{\gamma})\dfrac{\mathbf{S}\cdot\mathbf{L}_\gamma}{\hbar^2}|js_j,\gamma\rangle \nonumber \\
&=&
\sum_{\substack{k\in\beta\\ s_k=s_j}}
\langle is_i,\beta|\hat{\lambda}_{\beta}(r_{\beta})\dfrac{\mathbf{S}\cdot\mathbf{L}_\beta}{\hbar^2}|ks_k,\beta\rangle 
S_{kj}(\beta,\gamma)  \nonumber\\
&+&\sum_{\substack{k\in\gamma\\ s_i=s_k}}
S_{ik}(\beta,\gamma)
\langle ks_k,\gamma|\hat{\lambda}_{\gamma}(r_{\gamma})\dfrac{\mathbf{S}\cdot\mathbf{L}_\gamma}{\hbar^2}|js_j\gamma\rangle.
\label{inter-2}
\end{eqnarray}

Then, substituting  Eq. (\ref{intra}) into Eq. (\ref{inter-2}) we obtain for the two-center {\it inter}-atomic matrix elements
\begin{equation}
\langle is_i,\beta|H_{SO}^{inter}|js_j,\gamma\rangle
=
\sum_{\substack{k\in\beta\\ s_k=s_j\\ l_k=l_i}} S_{kj}(\beta,\gamma)\langle is_i|H_{SO}^{intra}|ks_k\rangle_\beta
+
\sum_{\substack{k\in\gamma\\ s_k=s_i\\ l_k=l_j}} S_{ik}(\beta,\gamma) \langle ks_k|H_{SO}^{intra}|js_j\rangle_\gamma.
\label{eq:1_7}
\end{equation}

Before concluding, we also use the Mulliken approximation in the {\it intra}-atomic two-center terms below 
\begin{eqnarray}
\langle is_i,\beta|  \sum_\alpha  \left(\hat{\lambda}_{\alpha}(r_{\alpha})\dfrac{\mathbf{S}\cdot\mathbf{L}_{\alpha}}{\hbar^2}\right) |js_j,\beta\rangle
=
\sum_\alpha \sum_{k,l\in\alpha} S_{ik}(\beta,\alpha) S_{lj}(\mathbf\alpha,\beta)
\langle ks_k|H_{SO}^{intra}|ls_l\rangle_\alpha,
\label{2in1}
\end{eqnarray}
recalling that $S_{ij}=\delta_{ij}$ when $\alpha=\beta$.

Finally, substituting Eq. (\ref{2in1}) and Eq. ({\ref{eq:1_7}) into Eq. (\ref{eq:1_3}), we obtain the expression for the matrix elements of spin-orbit interaction 
in the eH-TB framework 
\begin{eqnarray}                                                                                                                                                        
&\ &\langle is_i,\beta|H_{SO}|js_j,\gamma\rangle =
\delta_{\beta,\gamma} \sum_\alpha \sum_{k,l\in\alpha} S_{ik}(\mathbf{r}_\beta,\mathbf{r}_\alpha) S_{lj}(\mathbf{r}_\alpha,\mathbf{r}_\beta)
\langle ks_k|H_{SO}^{intra}|ls_l\rangle_\alpha + 
\label{HsoTB}\\
&\ &\left(1-\delta_{\beta,\gamma}\right) 
\left\{ 
\sum_{k\in\beta} S_{kj}(\mathbf{r}_\beta,\mathbf{r}_\gamma)\langle is_i|H_{SO}^{intra}|ks_k\rangle_\beta 
+
\sum_{k\in\gamma} S_{ik}(\mathbf{r}_\beta,\mathbf{r}_\gamma) \langle ks_k|H_{SO}^{intra}|js_j\rangle_\gamma
\nonumber
\right\}.
\end{eqnarray}
Notice that  the overlap matrix is block diagonal in the Hilbert space of the spinor.

\section{Spin-dependent Probability Current}

To obtain the spin-dependent probability current, we start with the single-particle time-dependent Schr\"odinger equation (TDSE) 
\begin{equation}
i\hbar\frac{\partial}{\partial t}\Psi(\mathbf{r},t) = 
\left\{-\frac{\hbar^2}{2m}\nabla^2  + V(\mathbf{r})  + 
\frac{1}{2(mc)^2}\mathbf{S}\cdot\boldsymbol{\nabla} V \times\mathbf{p}\right\}\Psi(\mathbf{r},t) ,
\label{TDSE}
\end{equation}
where $\mathbf{p} = -i\hbar \boldsymbol{\nabla}$ is the linear momentum operator and $\mathbf{S}$ is the spin operator in the basis of $S_z$ eigenstantes, 
\begin{eqnarray}
\mathbf{S} = \frac{\hbar}{2} 
\begin{bmatrix}
\hat{\mathbf{z}} & \hat{\mathbf{x}}-i\hat{\mathbf{y}} \\
\hat{\mathbf{x}}+i\hat{\mathbf{y}}& -\hat{\mathbf{z}}
\end{bmatrix},
\label{spinvec}
\end{eqnarray}
which acts on the spinor $\Psi(\mathbf{r},t) = \sum_{s=\uparrow,\downarrow} \Psi_s(\mathbf{r},t)$.

In the  framework of the eH-TB formalism,  we have
\begin{equation}
i\hbar\frac{\partial}{\partial t}\Psi(\mathbf{r},t) = \left\{ H_0 + H_{SO}\right\}\Psi(\mathbf{r},t),
\label{tb}
\end{equation}
with $H_0$ representing the eH-TB hamiltonian and $H_{SO}$ is given by Eq. (\ref{HsoTB}).

It has been pointed out that the conventional spin current, usually defined as 
$ \mathbf{j}_s = \dfrac{\hbar}{m} \Im m \left[ \Psi^*_s \boldsymbol{\nabla} \Psi_s \right]$, or 
alternatively in terms of the operator $(1/2)(\hat{\mathbf{v}}\hat{\mathbf{S}}+\hat{\mathbf{S}}\hat{\mathbf{v}})$,
is incomplete and unphysical under spin-flip hamiltonians.\cite{J-spinPRL,J-spinPRB,J-spinJapan,AJP-spin}
To obtain the expression for spin-depdendent continuity equation we base our approach on the work of Hodge et al.\cite{AJP-spin} 
Starting with Eq. (\ref{TDSE}), we multiply this equation by the complex conjugate of 
the wavefunction, $\Psi^*(\mathbf{r},t)$. Then conjugate Eq. (\ref{TDSE}) and multiply 
it by $\Psi(\mathbf{r},t)$. 
Subtracting the two equations, we obtain
\begin{equation}
\begin{split}
\sum_{s=\uparrow,\downarrow}\frac{\partial |\Psi_s|^2}{\partial t} = 
&-\sum_{s=\uparrow,\downarrow} \left\{ \boldsymbol{\nabla}\cdot\dfrac{\hbar}{2im}
\left(\Psi^*_s \boldsymbol{\nabla} \Psi_s -  \Psi_s \boldsymbol{\nabla} \Psi^*_s \right)\right\} \\
 &- \dfrac{1}{2m^2c^2}\sum_{s,s'} 
\left[ \Psi^*_s \mathbf{S}_{s,s'}\cdot\boldsymbol{\nabla} V\times\boldsymbol{\nabla} \Psi_{s'}
+
  \Psi_s \mathbf{S}^*_{s,s'}\cdot \boldsymbol{\nabla}V\times\boldsymbol{\nabla} \Psi^*_{s'}
\right]
\end{split} 
\label{cont1}
\end{equation}

\begin{equation}
\begin{split}
\sum_{s=\uparrow,\downarrow}\frac{\partial |\Psi_s|^2}{\partial t} &= 
-\nabla \cdot\left\{\dfrac{\hbar}{m}\sum_{\sigma=\uparrow,\downarrow} \Im m
\left[ \Psi^*_\sigma \nabla \Psi_\sigma \right] \right\} \\
 &+ \dfrac{1}{2m^2c^2}
\sum_{s,s'} 
\left[ \Psi^*_s \mathbf{S}_{s,s'}\cdot\boldsymbol{\nabla}\Psi_{s'}
+
\Psi_{s'} \mathbf{S}_{s,s'}\cdot\boldsymbol{\nabla}\Psi^*_{s}\right]\times
\boldsymbol{\nabla} V,
\end{split} 
\label{cont2}
\end{equation}
where we have used the property $\mathbf{S}^*_{s,s'} = \mathbf{S}_{s',s}$. Using the vector identity 
$\mathbf{A}\cdot(\mathbf{B}\times\mathbf{C}) = \mathbf{B}\cdot(\mathbf{C}\times\mathbf{A})$ on the second term of the RHS of Eq. (\ref{cont2}) we get
\begin{equation}
\begin{split}
\sum_{s=\uparrow,\downarrow}\frac{\partial |\Psi_s|^2}{\partial t} = 
&-\nabla \cdot\left\{\dfrac{\hbar}{m}\sum_{\sigma=\uparrow,\downarrow} \Im m
\left[ \Psi^*_\sigma \nabla \Psi_\sigma \right] \right\} \\
 &+ \dfrac{1}{2m^2c^2}
\sum_{s,s'} 
\boldsymbol{\nabla}(\Psi^*_s\Psi_{s'})\cdot\left(\boldsymbol{\nabla} V\times\mathbf{S}_{s,s'}\right),
\end{split} 
\label{cont3}
\end{equation}
and
\begin{eqnarray}
\boldsymbol{\nabla}(\Psi^*_s\Psi_{s'})\cdot\left(\boldsymbol{\nabla} V\times\mathbf{S}_{s,s'}\right)=
\boldsymbol{\nabla}\cdot\left(\Psi^*_s\Psi_{s'} \boldsymbol{\nabla} V\times\mathbf{S}_{s,s'} \right) 
-\Psi^*_s\Psi_{s'}\boldsymbol{\nabla}\cdot\left(\boldsymbol{\nabla}V\times\mathbf{S}_{s,s'}\right).
\label{cont3.1}
\end{eqnarray}
Then, by noting that $\boldsymbol{\nabla}\cdot\left(\boldsymbol{\nabla}V\times\mathbf{S}_{s,s'}\right) = 
\mathbf{S}_{s,s'}\cdot(\boldsymbol{\nabla}\times\boldsymbol{\nabla} V) 
- \boldsymbol{\nabla} V\cdot(\boldsymbol{\nabla}\times\mathbf{S}_{s,s'}) \equiv 0$ in Eq. (\ref{cont3.1}),
the continuity equation for the probability density  becomes \cite{AJP-spin}
\begin{equation}
\begin{split}
\sum_{s=\uparrow,\downarrow}\frac{\partial |\Psi_s|^2}{\partial t} = 
&-\boldsymbol{\nabla} \cdot\left\{\dfrac{\hbar}{m}\sum_{s=\uparrow,\downarrow} \Im m
\left[ \Psi^*_s \boldsymbol{\nabla} \Psi_s \right] \right\}\\
 &- \boldsymbol{\nabla}\cdot\left\{\dfrac{1}{2(mc)^2}\sum_{s,s'} 
\left[ \Psi^*_s \mathbf{S}_{s,s'} \Psi_{s'}
\right] \times \boldsymbol{\nabla} V\right\}~,
\label{cont4}
\end{split} 
\end{equation}
where we identify the conventional probability density current of well defined spin channel
\begin{equation}
\mathbf{j}_s = 
\dfrac{\hbar}{m} \Im m
\left[ \Psi^*_s \boldsymbol{\nabla} \Psi_s \right] 
\label{js}
\end{equation}
and the spin-mixed probability density current that is due to the SO interaction
\begin{equation}
\sum_{s'} \mathbf{j}^{_{SO}}_{s,s'} = 
\sum_{s'} \dfrac{1}{2(mc)^2} 
\left[ \Psi^*_s \mathbf{S}_{s,s'} \Psi_{s'}
\right] \times \boldsymbol{\nabla} V(\mathbf{r}).
\label{jso}
\end{equation}
The later has been associated with the torque dipole density\cite{J-spinPRL} or the angular spin current density.\cite{J-spinPRB}

Equations (\ref{js}) and (\ref{jso}) were derived for the continuum space representation, starting from the hamiltonian given in Eq. (\ref{TDSE}). 
To be consistent with the TB formalism that is used to calculate the wavepacket dynamics,  we need to use an effective mass $m^*$ instead of the free electron mass
$m_e$. In our calculations the internal consistency is obtained with $m^* = 0.42 m_e$, as described in the main text of the paper.

Furthermore, in order to have consistency between the TB matrix element  of Eq. (\ref{eq:1_4}) and                                   
the spin-orbit hamiltonian of Eq. (\ref{Hso}) we assume that the SOC constant is given by a Coulomb potential that is generated by an effective charge
$Q^\prime_\alpha$ -- to be determined ahead -- for each of the atomic species  in the double-helix.
That is
\begin{equation}
\boldsymbol{\nabla} V(\mathbf{r}) \approx \sum_\alpha \boldsymbol{\nabla} V_\alpha(\mathbf{r}-\mathbf{R}_\alpha)
\approx \sum_\alpha 
\dfrac{-e}{r_\alpha} 
\dfrac{d}{dr_\alpha}
\left(\dfrac{Q^\prime_\alpha}{r_\alpha}\right)\mathbf{r}_\alpha~.
\end{equation}
Therefore, Eq. (\ref{jso}) becomes
\begin{equation}
\sum_{s,s'} \mathbf{j}^{_{SO}}_{s,s'} \approx 
\dfrac{e}{2(mc)^2}
\sum_\alpha\dfrac{Q^\prime_\alpha}{r_\alpha^3}
\left\{
\sum_{s,s'} 
\left[ \Psi^*_s \mathbf{S}_{s,s'} \Psi_{s'}
\right] \times \mathbf{r}_\alpha
\right\}.
\label{jso1}
\end{equation}
Using the definition of $\mathbf{S}$, as given in Eq. (\ref{spinvec}), the term inside brackets can be expanded as
\begin{eqnarray}
&\ &\sum_{s,s'} \left[ \Psi^*_s \mathbf{S}_{s,s'} \Psi_{s'} \right] \times \mathbf{r}_\alpha \nonumber
= \\
&\ &\psi_\uparrow^*\psi_{\uparrow}(S_{\uparrow\uparrow}\times \mathbf{r}_\alpha)
+\psi_{\uparrow}^*\psi_{\downarrow}(S_{\uparrow\downarrow}\times \mathbf{r}_\alpha)
+\psi_{\downarrow}^*\psi_{\uparrow}(S_{\downarrow\uparrow}\times \mathbf{r}_\alpha)
+\psi_{\downarrow}^*\psi_{\downarrow}(S_{\downarrow\downarrow}\times \mathbf{r}_\alpha)=\nonumber\\
&\ &\frac{\hbar}{2}\left[\psi_\uparrow^*\psi_{\uparrow}(\hat{\mathbf{z}}\times\mathbf{r}_\alpha)
+\psi_{\uparrow}^*\psi_{\downarrow}((\hat{\mathbf{x}}-i\hat{\mathbf{y}})\times\mathbf{r}_\alpha)
+\psi_{\downarrow}^*\psi_{\uparrow}((\hat{\mathbf{x}}+i\hat{\mathbf{y}})\times \mathbf{r}_\alpha)
-\psi_{\downarrow}^*\psi_{\downarrow}(\hat{\mathbf{z}}\times\mathbf{r}_\alpha)\right] = \nonumber \\
&\ &\frac{\hbar}{2}\left[(\psi_\uparrow^*\psi_{\uparrow}-\psi_\downarrow^*\psi_{\downarrow})(\hat{\mathbf{z}}\times\mathbf{r}_\alpha)
+2\Re e(\psi_{\uparrow}^*\psi_{\downarrow})(\hat{\mathbf{x}}\times\mathbf{r}_\alpha)
+2\Im m(\psi_{\uparrow}^*\psi_{\downarrow})(\hat{\mathbf{y}}\times\mathbf{r}_\alpha)\right].
\label{eq_1.14}
\end{eqnarray}
Because the double-helix is aligned along the $\hat{\mathbf{z}}$ direction, we disregard the first term of Eq. (\ref{eq_1.14}), since it yields a probability 
density current perpendicular to that direction. Thus, we remain with
\begin{eqnarray}
\sum_{s,s'} \left[ \Psi^*_s \mathbf{S}_{s,s'} \Psi_{s'} \right] \times \mathbf{r}_\alpha \nonumber
&=& \frac{\hbar}{2}\left[
2\Re e(\psi_{\uparrow}^*\psi_{\downarrow})(\hat{\mathbf{x}}\times\mathbf{r}_\alpha)
+2\Im m(\psi_{\uparrow}^*\psi_{\downarrow})(\hat{\mathbf{y}}\times\mathbf{r}_\alpha)\right] \nonumber\\
&=& \hbar\left[
\Re e(\psi_{\uparrow}^*\psi_{\downarrow})(y_\alpha\hat{\mathbf{z}}-z_\alpha\hat{\mathbf{y}})
+\Im m(\psi_{\uparrow}^*\psi_{\downarrow})(z_\alpha\hat{\mathbf{x}}-x_\alpha\hat{\mathbf{z}})\right].
\label{eq_1.15}
\end{eqnarray}
Disregarding, again,  the terms perpendicular to the axis of the double-helix ($\hat{\mathbf{z}}$) we obtain for Eq. (\ref{jso1})
\begin{equation}
\sum_{s,s'} \mathbf{j}^{_{SO}}_{s,s'} 
\approx 
\dfrac{e\hbar}{2(m^*c)^2}
\left(\Re e(\psi_{\uparrow}^*\psi_{\downarrow})\sum_\alpha\dfrac{Q^\prime_\alpha y_\alpha}{r_\alpha^3}
-\Im m(\psi_{\uparrow}^*\psi_{\downarrow})\sum_\alpha\dfrac{Q^\prime_\alpha x_\alpha}{r_\alpha^3}
\right)\hat{\mathbf{z}}.
\label{jso2}
\end{equation}

Finally, it is necessary to determine the effective charges $Q^\prime_\alpha$ within the eH-TB formalism. To do so we turn 
to Eq. (\ref{radial}) and make use of the spin-orbit coupling parameters of Table \ref{SOCC}, so that
\begin{equation}
\langle \mathcal{R}_n(r_\alpha)| \dfrac{-e\hbar^2}{2(m^*c)^2}
\frac{1}{r_\alpha}\frac{d}{dr_\alpha}\left(\dfrac{Q^\prime_\alpha}{r_\alpha}\right) | \mathcal{R}_n(r_\alpha)\rangle 
= 
\lambda^{_{SOC}}_\alpha~,
\end{equation}
where $\mathcal{R}_n(r_\alpha)$ is the radial part of the STO for an atom with principal quantum number $n$, located at position $\mathbf{R}_\alpha$.
The integral can be solved as 
\begin{eqnarray}
\langle \mathcal{R}_n(r_\alpha)|
\frac{1}{|\mathbf{r} - \mathbf{R}_\alpha|^3} | \mathcal{R}_n(r_\alpha)\rangle 
&=& 
\dfrac{(2\zeta_{l,\alpha})^{2n_\alpha+1}}{(2n_\alpha)!}
\int_0^\infty\dfrac{|\mathbf{r}-\mathbf{R}_\alpha|^{2n_\alpha-2}e^{-2\zeta_{l,\alpha}|\mathbf{r}-\mathbf{R}_\alpha|}}{|\mathbf{r}-\mathbf{R}_\alpha|^3}r^2dr\nonumber \\
\nonumber \\
&=&
\dfrac{2\zeta_{l,\alpha}^3}{n_\alpha(n_\alpha-1)(2n_\alpha-1)},
\end{eqnarray}
for $n_\alpha > 1$, where $\zeta_{l,\alpha}$ is the orbital exponent of the STO with angular momentum $l$.
Since the spin-orbit coupling vanishes for $s$ orbitals, only angular momentum  $l=1$ is taken into consideration.
Thus, the effective charges $Q^\prime_\alpha$ are given by
\begin{equation}
Q^\prime_\alpha = \lambda^{_{SOC}}_\alpha \frac{(m^*c)^2}{e\hbar^2\zeta_{\alpha}^3} n_\alpha(n_\alpha-1)(2n_\alpha-1).
\end{equation}

\begin{figure}[htbp]
    \centering
 \includegraphics[width=0.5\linewidth]{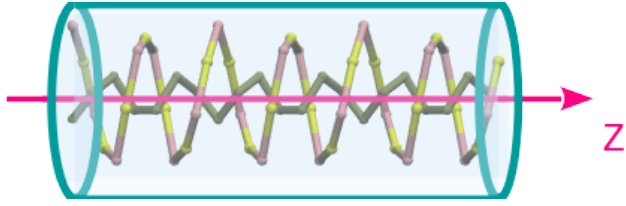}                             
 \caption{Portion of the SnIP double-helix along the $\hat{z}$ direction inside the  boundary surface $\mathcal{S}$ with inner volume $\mathcal{V}$.}
\label{volume}
\end{figure}

To calculate the electronic transport in the double helix we integrate the continuity equation over a volume $\mathcal{V}$ along the nanowire axis, as shown in Figure \ref{volume}, and apply the divergence theorem to obtain                                                                                                                 
\begin{equation}
\frac{\partial}{\partial t}  \left[ P_\uparrow + P_\downarrow \right]_\mathcal{_{V}} = 
-\sum_{s} \int_\mathcal{_{S}} {j}_{z,s} dA_{\perp}
-  \int_\mathcal{_{S}} 
\left[\sum_{s,s'} j^{_{SO}}_{s,s'}\right]_z dA_{\perp}~,
\label{detector}
\end{equation}
where $P_s$ is the spin-dependent electron population inside $\mathcal{V}$, associated with the spinor wavepackets $\Psi_s(\mathbf{r},t)$.
On the RHS, ${j}_{z,s} dA_{\perp}$ is the probability density flux with well defined spin projection across the surface caps perpendicular do the $\hat{z}$
direction. The last term accounts for the probability density flux of the mixed-spin current.

\section{Supporting Simulation Results}

\subsection{Spinless dynamics for M and P-SnIP}
It is instructive to consider the  spinless situation, when $H_{SO}$ is disregarded.
Panels a) and b), in Figure \ref{noSOC}, show the wavepacket dynamics in an M (left-handed) SnIP double-helix, and
panels c) and d) show the wavepacket dynamics in a P (right-handed) double-helix, both without SO interaction.
There is no observable difference for spinless wavepackets travelling along CW or CCW directions on M  or P-SnIP 
strands. 
The time-of-flight the wavepackets take to travel 45 unit cells ($\approx$ 35.7 nm) in the absence of SOC is $\tau_{tof}$ = 109 fs.

\begin{figure}[p]
    \centering
 \includegraphics[width=0.85\linewidth]{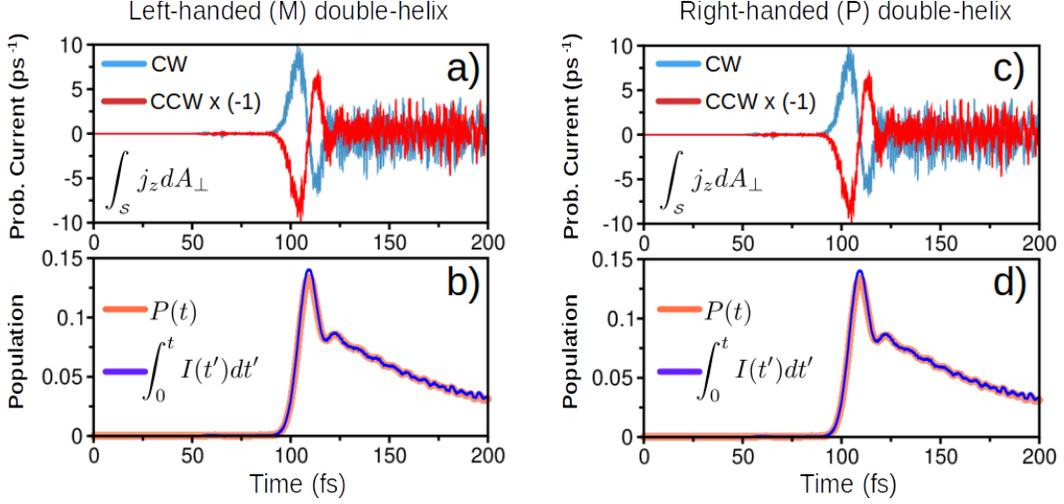}                             
 \caption{Wavepacket dynamics in an M (left-handed) SnIP double-helix without spin-orbit interaction. 
a) Probability current for the CW (blue) and CCW (red) spinless wavepackets; the later multiplied by the factor $-$1 to improve clarity.
b) Time-dependent electronic population in the detector volume $\mathcal{V}$, without the $j^{_{SO}}_{s,s'}$ term.
Analogous data for wavepacket dynamics in a P (right-handed) SnIP double-helix in panels c) and d). 
Ignoring the SOC, the propagation of spinless electrons for M-SnIP and P-SnIP strands are equal.}
\label{noSOC}
\end{figure}

\subsection{Spin-dependent dynamics for M and P-SnIP}

Figures \ref{SOC} to \ref{SOC2} describe the spin-dependent propagation dynamics for different combinations of the
spin angular momentum, direction of motion, and chirality of the underlying physical structure.

\begin{figure}[htbp]
    \centering
 \includegraphics[width=0.6\linewidth]{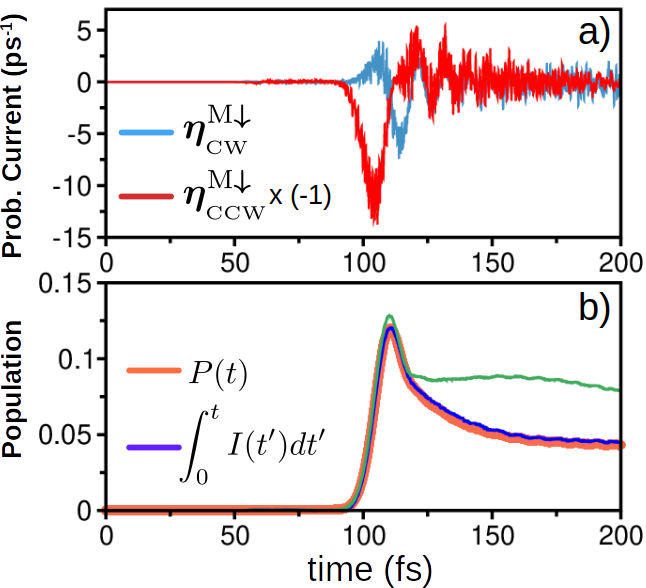}                             
 \caption{a) Probability density currents at the detector segment D produced by the counter-propagating wavepackets revealing the asymmetry of 
propagation velocity.              
Probability current for the $\eta^{M\downarrow}_{_{CW}} = \eta^+$ (blue) and $\eta^{M\downarrow}_{_{CCW}} =\eta^-$ (red) wavepackets, the later multiplied by the $-1$ 
factor for the sake of clarity. Note that $v(\eta^{M\downarrow}_{_{CCW}}) > v(\eta^{M\downarrow}_{_{CW}})$.
b) Time-dependent electronic population in the detector volume $\mathcal{V}$, as given by the LHS (orange) and the RHS (blue) terms of Eq. (19).
The green curve, obtained without the $\mathbf{j}^{_{SO}}_{s,s'}$ term, evinces the relevance of the spin-mixed component of $\mathbf{j}$.
}                                                                                                          
\label{SOC}
\end{figure}

\begin{figure}[htbp]
    \centering
 \includegraphics[width=0.6\linewidth]{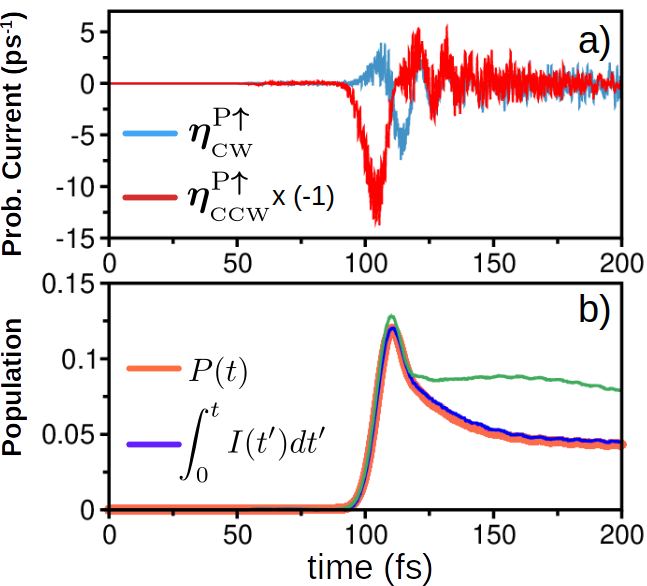}                             
 \caption{
a) Probability density currents at the detector segment D produced by the counter-propagating wavepackets revealing the asymmetry of 
propagation velocity.              
Probability current for the $\eta^{P\uparrow}_{_{CW}} = \eta^+$ (blue) and $\eta^{P\uparrow}_{_{CCW}} =\eta^-$ (red) wavepackets, the later multiplied by the $-1$ 
factor for the sake of clarity. Note that $v(\eta^{P\uparrow}_{_{CCW}}) > v(\eta^{P\uparrow}_{_{CW}})$.
b) Time-dependent electronic population in the detector volume $\mathcal{V}$, as given by the LHS (orange) and the RHS (blue) terms of Eq. (19).
The green curve, obtained without the $\mathbf{j}^{_{SO}}_{s,s'}$ term, evinces the relevance of the spin-mixed component of $\mathbf{j}$.
}                                                                                                          
\label{SOC1}
\end{figure}

\begin{figure}[htbp]
    \centering
 \includegraphics[width=0.6\linewidth]{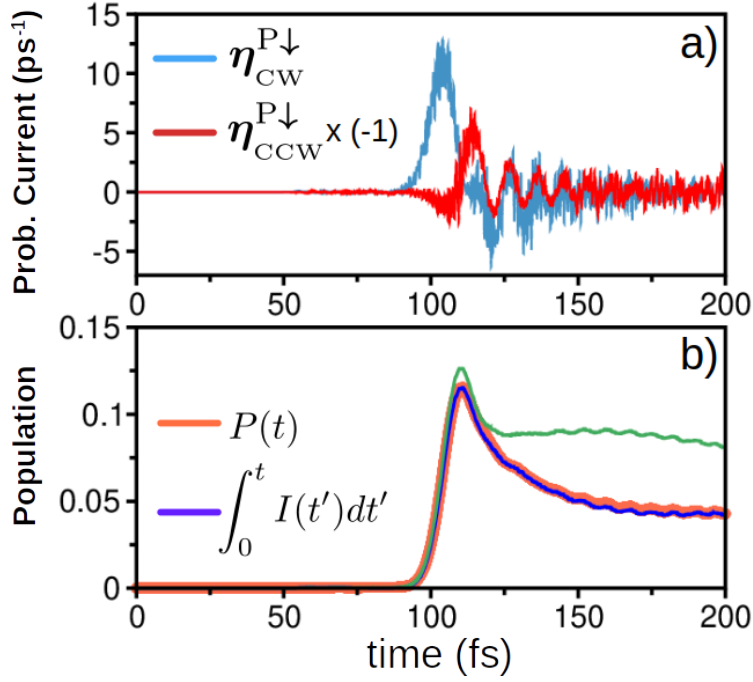}                             
 \caption{
a) Probability density currents at the detector segment D produced by the counter-propagating wavepackets revealing the asymmetry of 
propagation velocity.              
Probability current for the $\eta^{P\downarrow}_{_{CW}} = \eta^-$ (blue) and $\eta^{P\downarrow}_{_{CCW}} =\eta^+$ (red) wavepackets, the later multiplied by the $-1$ 
factor for the sake of clarity. Note that $v(\eta^{P\downarrow}_{_{CW}}) > v(\eta^{P\downarrow}_{_{CCW}})$.
b) Time-dependent electronic population in the detector volume $\mathcal{V}$, as given by the LHS (orange) and the RHS (blue) terms of Eq. (19).
The green curve, obtained without the $\mathbf{j}^{_{SO}}_{s,s'}$ term, evinces the relevance of the spin-mixed component of $\mathbf{j}$.
}                                                                                                          
\label{SOC2}
\end{figure}

\clearpage

\subsection{Spin-dependent dynamics for different propagation lengths in the M-SnIP double helix}

In relation to Figure \ref{lengths}, we consider the ring structure shown in Figure 4-a) of the paper, which is used to determine the propagation velocities 
of the CW and CCW wavepackets.                     
The structure accommodates two detector segments, D$_\text{CCW}$ and D$_\text{CW}$, both with the same size, but separated by a spacer segment of length 7.93 nm. 
The distance between the source (S) segment and each of the detectors (D$_\text{CW}$ and D$_\text{CCW}$) is varied as:
a) 7.9 nm (10 unit cells); b) 15.8 nm (20 unit cells); c) 23.8 nm (30 unit cells); and  d) 31.7 nm (40 unit cells).
Then, we apply the continuity equation, Eq. (\ref{detector}), to both detectors to evaluate the probability flux due to the counter propagating wavepackets.
The positive peaks indicate the inflow of charge in the detector whereas the negative peaks are associated with the outflow.
Table \ref{velocities} gives the time-of-flight for each case.

\begin{figure}[htbp]
    \centering
 \includegraphics[width=0.35\linewidth]{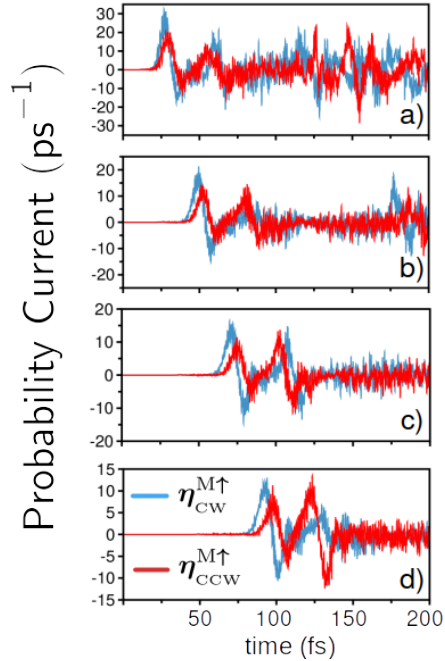}                             
 \caption{ Probability current for the $\eta^{M\uparrow}_{_{CW}} = -1$ (blue) and $\eta^{M\uparrow}_{_{CCW}} =+1$ (red) wavepackets,
evincing the asymmetry of propagation velocities: $v(\eta^{M\uparrow}_{_{CW}}) > v(\eta^{M\uparrow}_{_{CCW}})$.
The distance between the source (S) segment and each of the detectors (D$_\text{CW}$ and D$_\text{CCW}$) is varied as:
a) 7.9 nm (10 unit cells); b) 15.8 nm (20 unit cells); c) 23.8 nm (30 unit cells); and d) 31.7 nm (40 unit cells).
}                                                                                                          
\label{lengths}
\end{figure}

\begin{table*}
\caption{Time-of-flight ($\tau_{tof}$) as a functions of the distance between the source (S) segment and each of the detectors (D$_\text{CW}$ and D$_\text{CCW}$)} 
\begin{tabular}{c|c|c|c|c|c}
\hline \hline
 Distance between S & 7.9  & 15.8 & 23.8 & 31.7 & 35.7 \\
 and D segments     & (nm) & (nm) & (nm) & (nm) & (nm) \\
\hline
& & & & &\\
$\tau_{tof}$ (fs) & 27  & 50   & 71  &  93  &  103  \\ 
$\eta^{M\uparrow}_{_{\mathbf{CW}}}=\eta^- $  & & & & &\\
& & & & &\\
$\tau_{tof}$ (fs)  & 29  & 53   & 75  &  98  &  109  \\ 
$\eta^{M\uparrow}_{_{\mathbf{CCW}}}=\eta^+ $ & & & & &\\
& & & & &\\
\hline
\hline
\end{tabular}
\label{velocities} 
\end{table*}

A linear fit to the data of Table \ref{velocities} yields the ballistic propagation velocities $v(\eta^-) \approx$ 3.67 \AA/fs and $v(\eta^+) \approx$ 3.48 \AA/fs

\subsection{Comparison of Spinless and Spin-dependent dynamics}

Figure \ref{spin-nospin} presents a comparison between the spinless dynamics (without SO interaction) and the spin-dependent dynamics (with SOI considered).
For the sake of argument, we consider the M-SnIP double helix and the $\eta^{M\uparrow}_{_{CW}} = -1$ (blue) and $\eta^{M\uparrow}_{_{CCW}} =+1$ (red) wavepackets.
For the spinless case we simply have the CW (blue) and CCW (red) wavepackets. 
For presentation purposes, the spin-dependent probability currents have been shifted up and those for the spinless case have been shifted down.
The panels a), b), c) and d) correspond to 
different distances between the source (S) segment and each of the detectors (D$_\text{CW}$ and D$_\text{CCW}$), namely:
a) 7.9 nm (10 unit cells); b) 15.8 nm (20 unit cells); c) 23.8 nm (30 unit cells); and d) 31.7 nm (40 unit cells).
Figure \ref{spin-nospin} shows that the propagation velocity 
$v(\eta^-) \approx v_0$, where $v_0$ corresponds to the spinless case, whereas $v(\eta^+)$ decreases,
so that we have $v(\eta^-) \approx v_0 > v(\eta^+)$.
Simulations performed on the other wavepackets that comprise the Kramers doublets revealed the same behaviour.

\begin{figure}[htbp]
    \centering
 \includegraphics[width=1.00\linewidth]{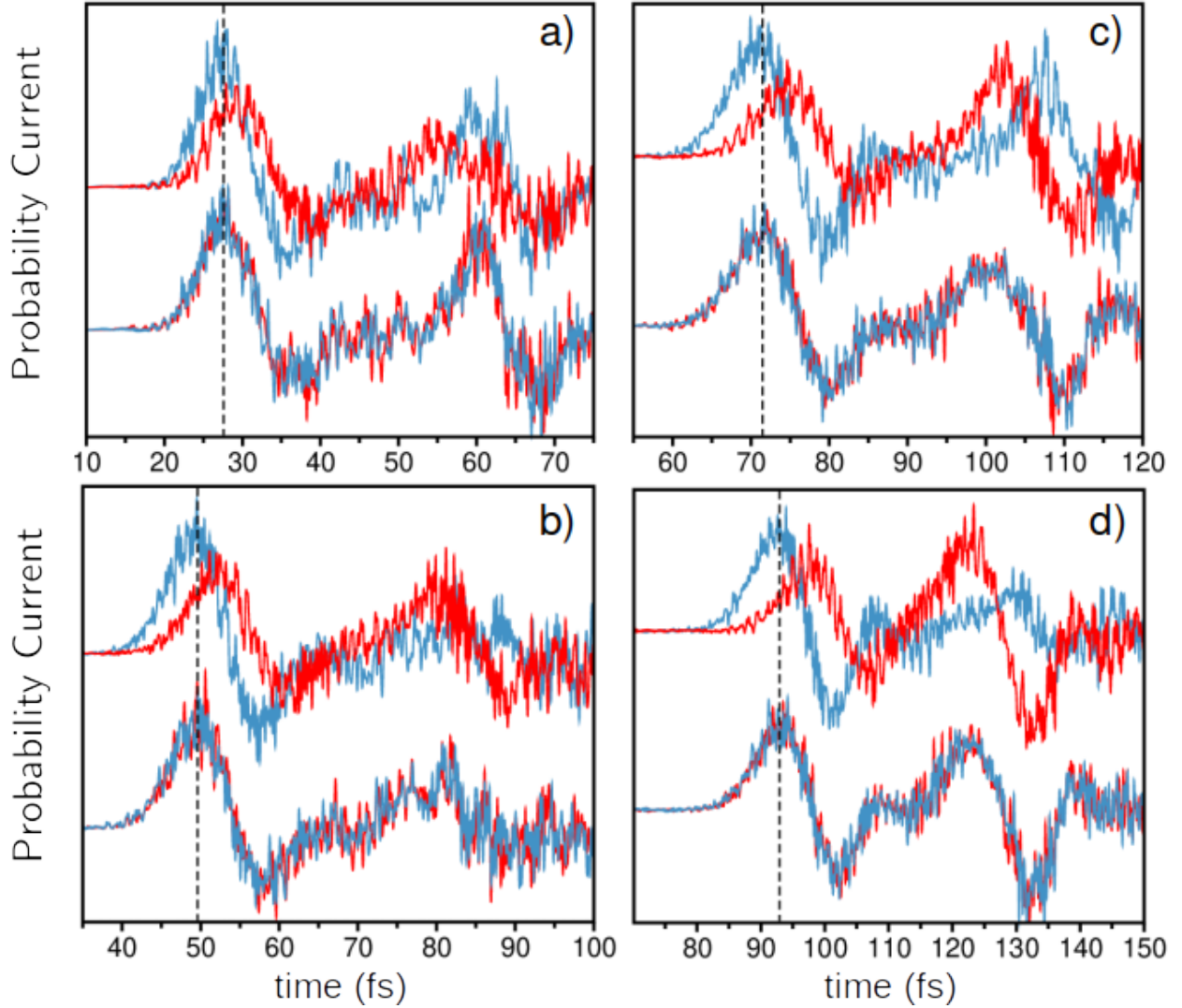}                             
 \caption{ Probability current for the $\eta^{M\uparrow}_{_{CW}} = -1$ (blue) and $\eta^{M\uparrow}_{_{CCW}} =+1$ (red) wavepackets,
shifted up. Probability current for the CW (blue) and CCW (red) spinless wavepackets, shifted down.
The distance between the source (S) segment and each of the detectors (D$_\text{CW}$ and D$_\text{CCW}$) is varied as:
a) 7.9 nm (10 unit cells); b) 15.8 nm (20 unit cells); c) 23.8 nm (30 unit cells); d) 31.7 nm (40 unit cells).
}                                                                                                          
\label{spin-nospin}
\end{figure}

\clearpage

\bibliography{myRefs}

\end{document}